\documentclass[a4,12pt,epsf]{article}
\usepackage{graphicx,amsmath}
\usepackage{amsmath,amsfonts,latexsym,amssymb}
\usepackage{verbatim,amsthm,graphics}

\usepackage{mathrsfs}
\def\ben{\begin{equation}}
\def\een{\end{equation}}
\def\bena{\begin{eqnarray}}
\def\eena{\end{eqnarray}}

\newcommand{\T}{{\mathbb T}}
\renewcommand{\S}{{\mathbb S}}
\newcommand{\mL}{{\mathcal L}}

\newcommand{\G}{{\mathcal G}}
\newcommand{\K}{{\mathcal T}}
\newcommand{\mr}{{\mathbb R}}

\newcommand{\mc}{{\mathbb C}}
\newcommand{\Z}{{\mathbb Z}}

\renewcommand{\H}{{\mathcal H}}

\newcommand{\e}{{\rm e}}
\renewcommand{\d}{{\rm d}}

\theoremstyle{definition}
\newtheorem{thm}{Theorem}

\newtheorem{lemma}{Lemma}
\newtheorem{proposition}{Proposition}

\begin{document}

%\begin{flushright}
%Sofia University\\
%\end{flushright}
%%%%%%%%%%%%%%%%%%%%%%%%%%%%%%%%%%%%%%%%%%%%%%%%%%%%%%%%%%%%%%%%%%%

\title{A uniqueness theorem for stationary Kaluza-Klein black holes}

\author{
Stefan Hollands$^{1}$\thanks{\tt HollandsS@Cardiff.ac.uk}\:,
%%%
Stoytcho Yazadjiev$^{2}$\thanks{\tt yazadj@theorie.physik.uni-goe.de}\:,
\\ \\
%%%
{\it ${}^{1}$ School of Mathematics, Cardiff University} \\
{\it Cardiff, CF24 4AG, UK} \medskip \\
%%%
%%%
{\it ${}^{2}$Department of Theoretical Physics, Faculty of Physics, Sofia
University} \\
{\it 5 J. Bourchier Blvd., Sofia 1164, Bulgaria} \\
    }

\date{5 May 2009}

\maketitle

\begin{abstract}
We prove a uniqueness theorem for stationary $D$-dimensional Kaluza-Klein black holes with
$D-2$ Killing fields, generating the symmetry group ${\mathbb R} \times U(1)^{D-3}$.
It is shown that the topology and metric of such black holes is uniquely determined by the angular momenta
and certain other invariants consisting of a number of real moduli, as well as integer vectors
subject to certain constraints.
%The statement and proof of the theorem is a generalization of that given in gr-qc/0707.2775 for five-dimensional asymptotically %flat black holes.
\end{abstract}

\section{Introduction}
\label{sec0}

The classic black hole uniqueness theorems state that four dimensional, stationary, asymptotically
flat black hole spacetimes are uniquely determined by their mass and angular momentum in the vacuum case,
and by their mass, angular momentum, and charge in the Einstein-Maxwell case. The solutions
are in fact given by the Kerr metrics in the first case and the Kerr-Newman metrics in the second.
  This was proven in a series of
papers~\cite{Carter,Robinson,Bunting,Mazur,Hawking,H72,Israel67}; for a coherent exposition clarifying many important
details and providing a set of consistent technical assumptions see~\cite{chrcos}.

The black hole uniqueness theorem is not true as stated in general spacetime dimensions $D \ge 5$. For
example, in $D=5$ dimensions, there exist asymptotically flat, stationary vacuum black holes with the same mass
and angular momenta, but with non-isometric spacetime metrics, and in fact even different topology~\cite{Myers,Emparan,senkov,elvang,chen,evslin,elvang1}.
One would nevertheless hope that a similar uniqueness theorem still applies if additional invariants (``parameters'')
are specified beyond the mass and angular momenta. Unfortunately, except in the static case~\cite{Gibbons,Rogatko},
such a classification result is not known, nor is it known what could be the nature of the additional invariants.

In this paper, we consider the special case of stationary
vacuum black hole spacetimes in dimension $D \ge 4$ with a compact, non-degenerate, connected horizon, admitting
$D-3$ commuting additional Killing fields with closed orbits. The spacetimes that we consider asymptote to a flat
Kaluza-Klein space with $1,2,3$ or $4$ large spatial dimensions and a corresponding number of toroidal extra
dimensions. We will first show how to associate certain invariants to such a spacetime consisting of
a collection of ``moduli'' $\{l_i \in \mr_{>0}\}$ and certain generalized ``winding numbers'' $\{\underline a_i \in \Z^{D-3}\}$.
The moduli may be thought of as the length of various rotation ``axis'' within the spacetime, whereas
the winding numbers characterize the nature of the action of the $D-3$ rotational symmetries near a given axis.
The collection of these winding numbers uniquely characterizes the topology and symmetry structure of
the exterior of the black hole, and we refer to it as the ``interval structure'' of the manifold.
This analysis also implies that the horizon must be topologically the
cartesian product a torus of the appropriate dimension and either a 3-sphere, ring ($\S^2 \times \S^1$), or
Lens-space $L(p,q)$.

Our notion of interval structure may be viewed as a generalization of what has been called ``weighted orbit space''
in the mathematics literature on 4-manifolds with torus action~\cite{orlik1,orlik2}, but the latter notion does not
involve the moduli $\{l_i\}$. Also, in the context of
stationary black holes, a similar notion called ``rod structure'' was first considered by~\cite{Harmark,Oelsen};
see~\cite{Reall} for the static case. The main difference between this and our notion is that
our winding numbers are found to obey an integrality condition as well as certain other constraints,
whereas there were no such constraints in~\cite{Harmark,Oelsen}. The latter are a necessary and sufficient condition
for the spacetime to have the
structure of a smooth manifold with torus action.
These topological considerations are described in detail in sec.~\ref{sec2}.

We will then prove a uniqueness theorem which states that there can be at most one black hole
spacetime with the same angular momenta and interval structure\footnote{It has been
brought to our attention that a conjecture in this direction had been made at the talk~\cite{Harmarktalk},
see also~\cite{Harmark}.}.
Our uniqueness theorem generalizes
a theorem proved in a previous paper~\cite{Hollands:2007aj} on asymptotically flat vacuum black holes in $D=5$ dimensions;
see also~\cite{Hollands:2007qf} for the Einstein-Maxwell case.

The proof of the theorem proceeds in two steps: First, one reduces the full Einstein equations to
equations on the space of symmetry orbits. Because the spacetime is assumed to have a total number of $D-2$
independent commuting Killing fields, the space of symmetry orbits is two-dimensional---in fact it is shown to
be a manifold with boundaries and corners homeomorphic to a half-plane. The parameters $\{l_i\}$ are essentially
the lengths of the various boundary segments. The arguments in the first step are topological in
nature, and the only role of Einstein's equations is to provide additional information about the fundamental group
of the manifold via the topological censorship theorem~\cite{Chrusciel:2008uz}. That information is needed to
rule out the presence of conical singularities in the orbit space\footnote{Here our analysis also fills a gap
in our previous paper~\cite{Hollands:2007aj}, where the absence of such conical singularities had to be assumed by hand.}.
Our results in this part may be thought of as a generalization of~\cite{orlik1,orlik2} to a higher dimensional situation.

The second step is to cast the reduced Einstein equations on the orbit
space into a suitable form. Here, we make use of a formulation due to~\cite{Maison} involving certain
potentials. The form of the equations leads to a partial differential equation for a quantity representing the
``difference'' between any two black hole metrics of the type considered which has been called ``Mazur identity''~\cite{Mazur}. Using this identity, one can prove the uniqueness theorem. The vectors $\{ \underline a_i \}$
and parameters $\{l_i\}$ are important
to treat the boundary conditions of the differential equation. The arguments in the second step are
geometrical/analytical, and involve the use of Einstein's equations in an essential way.
The simpler case of a spherical black hole with trivial interval structure was previously treated by a similar method in~\cite{Ida}.

While our uniqueness theorem in higher dimensions is in some ways similar to the corresponding theorem in four
dimensions, there are some notable differences. The first, more minor, difference is that higher dimensional black holes
are not only classified by the mass and angular momenta, but in addition depend on the interval structure. In
$D=4$ the interval structure of a single black hole spacetime is trivial.
A more substantial difference is that in $D=4$ dimensions, the additional axial Killing field is in fact
guaranteed by the rigidity theorem~\cite{Hawking,Moncrief,Friedrich,Racz,Chrusciel}. While a generalized
rigidity theorem can be established in $D$ dimensions~\cite{HIW,Moncrief:2008mr,Hollands:2008wn}, this theorem now only guarantees at least one additional axial Killing field. For the arguments of the present paper to work,
we need however $D-3$ commuting axial Killing fields. It does not seem likely that
our theorem covers all asymptotically Kaluza-Klein, stationary black hole spacetimes in $D$ dimensions.

A third difference is that we have not been able so far to establish for which given set of angular
momenta and interval structure there actually exists a regular black hole solution. The situation in this
regard is in fact unclear even in five asymptotically large dimensions with no small extra dimensions.
Here, solutions corresponding to various simple interval structures have been constructed.
These include solutions with horizon topology
$\S^3, \S^2 \times \S^1, L(p,q)$, which are the possible toplogies allowed by our uniqueness theorem. However,
by contrast with the cases $\S^3, \S^2 \times \S^1$~\cite{Myers,Emparan,senkov,elvang}, the black holes with lens space horizon topology found so far~\cite{chen,evslin} are not regular, and are thus actually not covered by our theorem. The situation is very different in four dimensions. Here the interval structure for single black hole spacetimes only involves the specification of a single parameter (related to the area of the horizon), and a regular black hole solution is known to exist for any choice of this parameter and the angular momentum---the corresponding Kerr solution. The mass, surface
gravity, angular velocity of the horizon etc. of the solution can all be expressed in terms of these parameters.

%%%%%%%%%%%%%%%%%%%%%%%%%%%%%%%%%%%%%%%%%%%%%%%%%%%%%%%%%%%%%%%%%%%

%\draft
\sloppy

\section{Description of the problem, assumptions, notations}
\label{sec1}
Let $(M,g)$ be a $D$-dimensional, stationary black hole spacetime satisfying the vacuum Einstein
equations, where $D \ge 4$. The asymptotically timelike Killing field is called $t$, so
$\pounds_t g = 0$. We assume that $M$ has $s+1$ asymptotically flat large spacetime dimensions
and $D-s-1$ asymptotically small extra dimensions, where $s>0$.
More precisely, we assume that a subset  of $M$ is diffeomorphic to the cartesian product of
$\mr^s$ with a ball removed---corresponding to the asymptotic region of the large spatial
dimensions---and $\mr \times \T^{D-s-1}$---corresponding to the time-direction and small dimensions.
We will refer to this region as the asymptotic region and call it $M_\infty$.
The metric is required to behave in this region like
\ben\label{standard}
g = -\d\tau^2 + \sum_{i=1}^s \d x_i^2 + \sum_{i=s+1}^{D-1} \d\varphi_i^2 + O(R^{-\alpha}) \, ,
\een
where $\alpha>0$ is some constant, and where $O(R^{-\alpha})$ stands for metric components that drop off faster than $R^{-\alpha}$ in the radial
coordinate $R = \sqrt{x_1^2+...+x_s^2}$, with $k$-th derivatives dropping off faster
than $R^{-\alpha-k}$. These terms are also required to be independent of the coordinate
$\tau$, which together with $x_i$ forms the standard  cartesian coordinates on $\mr^{s,1}$.
The remaining coordinates $\varphi_i$ are $2\pi$-periodic and parametrize the torus
$\T^{D-s-1}$. The timelike Killing field is assumed to be equal to $\partial/\partial \tau$ in
$M_\infty$. We call spacetimes satisfying these properties
``asymptotically Kaluza-Klein'' spacetimes\footnote{For the axisymmetric spacetimes considered in
this paper, we will derive below  a stonger asymptotic expansion, see
eq.~\eqref{metricinfty}.}.

The domain of outer communication is defined by
\ben
\langle \! \langle M \rangle \! \rangle =
I^+ \left(
M_\infty
\right) \cap
I^- \left(
M_\infty
\right)\,,
\een
where $I^\pm$ denote the chronological past/future of a set. The black
hole region $B$ is defined as the complement in $M$ of the causal past of the
asymptotic region, and its boundary $\partial B = H$ is called the (future)
event horizon.

In this paper, we also assume the existence of $D-3$ further linearly independent Killing fields,
$\psi_1, \dots, \psi_{D-3}$, so that the total number of Killing fields is
equal to the number of spacetime dimensions minus two. These are required to mutually
commute, to commute with $t$, and to have periodic orbits which close
for the first time after $2\pi$. The Killing fields $\psi_i$ are referred to
as ``axial'' by analogy to the four-dimensional case, even though their
zero-sets are generically higher dimensional surfaces rather than ``axis''
in $D>4$. We also assume that, in the asymptotic region $M_\infty$, the action of the axial symmetries
 is conjugate to the standard rotations in the cartesian product of
flat Minkowski spacetime $\mr^{s,1}$ times the standard flat torus $\T^{D-s-1}$.
In other words, $\psi_i = \partial/\partial \varphi_i$ or\footnote{The
notation $[x]$ means the largest integer $n$ such that $n \le x$.} $\psi_j = x_{2j-1} \partial_{x_{2j}}
- x_{2j} \partial_{x_{2j-1}}$ for $j=1, \dots, [s/2]$
in $M_\infty$. The group
of isometries is hence $\G = \mr \times \K$, where $\mr$ corresponds to the flow of
$t$, and where $\K = \T^{D-3}$ corresponds to the commuting flows of the axial
Killing fields. Looking at the action of $\G$ on the asymptotic region, it is evident
that an asymptotically Kaluza-Klein spacetime can have at most $[s/2]+D-s-1$ commuting
axial Killing fields. If this number is $D-3$ as we are assuming, then $s$ can be either $1,2,3$ or $4$. Thus our
spacetime is asymptotically the direct product of $2,3,4$- or $5$-dimensional Minkowski
spacetime and a $(D-2),(D-3),(D-4)$- or $(D-5)$-dimensional flat torus.

We are going to analyze the uniqueness properties of the asymptotically
Kaluza-Klein spacetimes just described. Unfortunately, in order to make our
arguments in a consistent way, we will have to make certain further technical
assumptions about the global nature of $(M,g)$ and the action of the symmetries.
Our assumptions are in parallel to those made by Chrusciel and Costa
in their study~\cite{chrcos} of 4-dimensional stationary black holes. The
requirements are (a) that $\langle \! \langle M \rangle \! \rangle$ contains an acausal, spacelike, connected
hypersurface $S$ asymptotic to the $\tau = 0$ surface in the asymptotic region $M_\infty$,
whose closure has as its boundary $\partial S = \H$ a cross section
of the horizon. We assume $\H$ to be compact and (for simplicity) to be connected.
(b) We assume that the orbits of $t$ are complete. (c) We assume that
the horizon is non-degenerate. (d) We assume that $\langle \! \langle M \rangle \! \rangle$
is globally hyperbolic.
%(e) We will assume that the action of $\K$ is such
%that there are no points with discrete isotropy subgroup in $\langle \! \langle M \rangle \! \rangle$.
%The last property follows automatically in $D=4$~\cite{chrcos}.

For the spacetimes described, one of the following two statements is true:  If $t$ is tangent to the
null generators of $H$ then the spacetime must be static~\cite{Sudarsky}.
On the other hand, if $t$ is not tangent to the
null generators of $H$, then the rigidity theorem~\cite{HIW} implies that there exists a
linear combination
\begin{equation}\label{Kdef}
K = t + \Omega_{1}^{} \psi_{1} + \dots +
\Omega_{D-3} \psi_{D-3}, \quad \Omega^{}_{i} \in \mr
\end{equation}
so that the Killing field $K$ is tangent and normal to the null
generators of the horizon $H$, and
\begin{equation}\label{orth}
g(K, \psi_{i}) = 0 \quad \text{on $H$.}
\end{equation}
From $K$, one may define the surface gravity of the black hole by
$\kappa^2 = \lim_H (\nabla_a f) \nabla^a f/f$, with $f=(\nabla^a K^b)
\nabla_a K_b$, and it may be shown that $\kappa$ is constant
on $H$~\cite{Waldbook}. In fact, the non-degeneracy condition implies
$\kappa > 0$.

In the first case, one can prove that the spacetime is actually unique~\cite{Israel67},
and in fact isometric to the Schwarzschild
spacetime when $D=4$,
for higher dimensions see~\cite{Gibbons,Rogatko}.
In this paper, we will be concerned with the second
case, and we will give a uniqueness theorem for such spacetimes.
Of particular importance for us will be the orbit space
$\hat M=\langle \! \langle M \rangle \! \rangle/\G$, so in the
next section we will look in detail at this space.

%%%%%%%%%%%%%%%%%%%%%%%%%%%%%%%%%%%%%%%%%%%%%%%%%%%%%%%%%%%%%%%%%%%%%

\section{Analysis of the orbit space}
\label{sec2}

\subsection{Manifolds with torus actions}
To begin, we consider
a somewhat simpler situation, namely an orientable, smooth, compact connected Riemannian manifold $\Sigma$
of dimension $s \ge 3$, with a smooth effective\footnote{This means that if $k \cdot x = x$ for all $x \in \Sigma$,
then $k$ is necessarily the identity. Given an action of the above type, one may always pass to an effective action by taking
a quotient of $\K$ if necessary.} action of the $N$-dimensional torus $\K=\T^N$. Thus,
we assume that ${\rm Diff}(\Sigma)$ contains a copy of $\K$. Such actions have been
analyzed and classified in the case $s=4$ in a classic work by Orlik and Raymond~\cite{orlik1, orlik2}, and---repeating
many of their arguments---in~\cite{Hollands:2007aj}. Some of our arguments for general $s$ are in parallel
with this case, others are not.

We may equip $\Sigma$ with a Riemannian
metric $h$, and by averaging $h$ with the action of $\K$ if necessary, we may assume that $\K$ acts by
isometries of $h$. Later, $\Sigma$ will be a spatial slice of our physical spacetime (so that $s=D-1$) and
$N$ will be taken to be $D-3$, but for the moment this is not relevant.

It will be useful to view the $N$-torus as the quotient $\mr^N/\Lambda_N$, where
$\Lambda_N = (2\pi\Z)^N$ is the standard $2\pi$-periodic $N$-dimensional lattice. Elements $k \in \K$ will
consequently be identified with equivalence classes of $N$-dimensional vectors,
$k=[\tau_1, \dots, \tau_{N}] \in \mr^N/\Lambda_N$. The standard basis of $\Lambda_N$ will be denoted
$\underline b_1, \dots, \underline b_N$, i.e.,
$$
\underline b_i = (0, \dots, 0, 2\pi, 0, \dots, 0) \, ,
$$
where the non-zero entry is in the $i$-th position.
Various facts about lattices that we will use in this section may be found in the
classic monograph~\cite{cassels}.

We denote the commuting Killing fields
generating the action of $\K$ by $\psi_i, i=1, \dots, N$. The flows of these vector fields are
denoted $F_i^\tau$, and we assume that they are normalized so that the flows are periodic with
period $2\pi$, so $F^{2\pi}_i(x) = x$ for any $x \in \Sigma$, and any $i$. The action of a group
element $k = [\tau_1, \dots, \tau_N]$ on a point is abbreviated by
\ben
k \cdot x = F^{\tau_1}_1 \circ \dots \circ F^{\tau_{N}}_{N}(x) \, .
\een
We also abbreviate the action of $k$ on a tensor field $T=T_{a_1 \dots a_q}{}^{b_1\dots b_r}$ on $\Sigma$
by $k \cdot T = [F^{\tau_1}_1 \circ \dots \circ F^{\tau_{N}}_{N}]^* T$, where the $*$ denotes the
push-forward/pull-back of the tensor field. Because the Killing fields commute, we have in particular
$k \cdot \psi_i = \psi_i$ for any $k\in \K$. If $\psi_1, \dots, \psi_{N}$ are Killing fields as above,
then so are $\hat \psi_1, \dots,
\hat \psi_{N}$, where
\ben\label{hatpsi}
\hat \psi_i = \sum_{i=1}^N A_{ij} \psi_j\, , \quad
\pm A = \pm \left(
\begin{matrix}
A_{11} & \dots & A_{N \, 1}\\
\vdots &       & \vdots\\
A_{1 \, N} & \dots & A_{N \, N}
\end{matrix}
\right) \in SL(N, \Z) \, .
\een
Another way of saying this is that we may conjugate the action of $\K=\T^{N}$
by the inner automorphism\footnote{The automorphism property is
$\alpha_A(kk') = \alpha_A(k)\alpha_A(k')$ for all $k,k' \in \K$.}
$\alpha_A([\underline{\tau}]) = [\underline{\tau} A^T]$ of $\K$, and the modified
Killing fields $\hat \psi_i$ generate the conjugated action.
The freedom of choosing different $2\pi$-periodic Killing fields to generate the
action of $\K=\mr^N/\Lambda_N$ is closely related to possibility of choosing different bases in
the lattice $\Lambda_N$, because any such change of basis is implemented by an integer matrix $A$
with $\det A = \pm 1$.

As is standard, we define the orbit and the isotropy subgroup associated with a point by, respectively
\ben
O_x = \{ k \cdot x \mid k \in \K\}\, , \quad
I_x = \{ k \in \K \mid k \cdot x = x \} \, .
\een
$I_x$ is a closed (hence compact) subgroup of $\K$, and
$O_x$ is a smooth manifold that can be identified with $\K/I_x$. Being compact and abelian,
$I_x$ must be isomorphic to $\T^n \times \prod \Z/p_j \Z$. A more precise description of the action
$I_x$ in an open neighborhood of $x$ will be given below. The set of all orbits $\hat \Sigma = \{O_x \mid x \in \Sigma\}$
is called the factor space and is also written as $\hat \Sigma = \Sigma/\K$. It is not a manifold for
general group actions.

It will be useful to define the non-negative, symmetric $N \times N$
Gram matrix of the Killing fields,
\ben
f_{ij} = h(\psi_i, \psi_j) \, .
\een
It will also be convenient to distinguish points in $\Sigma$ according to the dimension of their
orbit. For this, we define
\bena\label{srdef}
S_r &=& \{ x \in \Sigma \mid \dim O_x = r\}\nonumber\\
&=& \{ x \in \Sigma \mid {\rm rank}[f(x)] = r\}\nonumber\\
&=& \{ x \in \Sigma \mid {\rm dim} \, I_x = n = N-r\} \, .
\eena
Evidently, $n=N-r$ is also equal to the number of independent linear combinations
of the Killing fields $\psi_1, \dots, \psi_{N}$ that vanish at points of $S_r$.
Clearly, we have
\ben\label{decompdef}
\Sigma = \bigcup_{r=0}^{N} S_r \, .
\een
\begin{lemma}\label{lemma1}
Let $(\Sigma, h)$ be a Riemannian manifold of dimension $s$, with
$N$ mutually commuting Killing fields $\psi_i, i=1, \dots, N$. Let
$f_{ij}$ be the Gram matrix, and let $x$ be a point such that
${\rm rank}[f(x)] = r$. Then it follows that
$N-r \le [(s-r)/2]$.
\end{lemma}
{\em Proof:}
Let $V_x \subset T_x \Sigma$ be the span of the Killing fields $\psi_i |_x , i=1, \dots, N$ at
$x$, and let $W_x$ be the orthogonal complement. The assumptions of the lemma mean
that the dimension of $V_x$ is $r$, and that there exist $N-r$ linear combinations
of $\psi_i |_x , i=1, \dots, N$ that vanish. By forming suitable linear combinations of the Killing
fields, we may hence assume that ${\rm span}\{\psi_i |_x, i=1, \dots, r \} = V_x$, and that
$\psi_i |_x = 0, i=r+1, \dots, N$. Let $D$ be the derivative operator
of $h$, and let $t_i = D \psi_i |_x$, where $i=r+1, \dots, N$. Then each $t_i$
is a linear map $t_i: T_x \Sigma \to T_x \Sigma$. The Killing equation implies that
$t_i$ is skew symmetric with respect to the bilinear form $h: T_x \Sigma \times T_x \Sigma \to \mr$,
i.e. $h(t_i X, Y) = -h(X, t_i Y)$. Evaluating the $D$-derivative of the commutator $[\psi_i, \psi_j] = 0$ at
$x$ for $r<i,j\le N$ then implies that the corresponding commutator
$t_i t_j - t_j t_i=0$ vanishes, too. Evaluating the derivative of the commutator $[\psi_i, \psi_j] = 0$ at $x$ for
$r<i\le N$ and $0<j\le r$ then furthermore shows that $t_i \restriction V_x = V_x$, and
consequently $t_i \restriction W_x = W_x$. Now let us choose an orthogonal basis $\{e_1, \dots, e_{s-r}\}$
of $W_x$, and use that to identify $t_i, r<i\le N$ with a linear map $\mr^{s-r} \to \mr^{s-r}$.
These linear maps must hence skew symmetric, i.e., commuting elements of the Lie-algebra
${\mathfrak o}(s-r, \mr)$. They must also be linearly independent. Indeed, assume on the
contrary that that a non-trival linear combination $\lambda_1 t_{r+1}+\dots+\lambda_{N-r} t_{N}$
vanishes. Then both the Killing field $s = \lambda_1 \psi_{r+1} +\dots+\lambda_{N-r} \psi_{N}$,
as well as its derivative $Ds$ vanish at the point $x$. It is a well-known property of
Killing fields (see e.g.~\cite{Waldbook}) that a Killing field vanishes identically
on a connected Riemannian manifold if it vanishes at a point together with its derivative.
Hence, the Killing fields $\psi_i, r < i \le N$ would be linearly dependent, a contradiction.
Thus, we conclude that the linear maps $t_i, r< i \le N$ may be viewed as forming a $(N-r)$-dimensional
abelian subalgebra of ${\mathfrak o}(s-r, \mr)$.
Any maximal abelian subalgebra of ${\frak o}(s-r, \mr)$
has dimension $[(s-r)/2]$, so $N-r \le [(s-r)/2]$.
\qed

In the situation considered later in this section, we have $N=s-2$ Killing fields.
The lemma then implies that the sets $S_r$ are non-empty only for
$r=s-2, s-3, s-4$, so we have $\Sigma = S_{s-2} \cup S_{s-3} \cup S_{s-4}$.

\medskip

Our task will now be to construct, for each orbit $O_x$, an open neighborhood of it
and a coordinate system in which we can explicitly understand the action of the group $\K$.
We will then be able to locally take the quotient of this neighborhood by $\K$ and
thereby get a local description of the orbit space. By patching the local regions together,
we will be able to characterize the manifold structure of the orbit space.

Let $x$ be an arbitrary but fixed
point in $S_r$. Then the dimension of $O_x$ is $r$, and the dimension of the isotropy group $I_x$
is $n=N-r$. As we have just seen, $n$ may only take on the values $0,1,\dots,[(s-r)/2]$.
%When $N=s-2$, then we will see that
%$O_x$ has a neighborhood in $\hat \Sigma$ locally isomorphic to the tip of
%a cone $\mr^2/(\Z/q \Z)$ for $n=0$, that $O_x$ has a neighborhood in $\hat \Sigma$ locally
%diffeomorphic to a half-space $\mr_{>0} \times \mr$ for $n=1$, and that
%$O_x$ has a neighborhood in $\hat \Sigma$ locally diffeomorphic to a corner
%$\mr_{>0} \times \mr_{>0}$ for $n=2$.
We first show that if $x \in S_r$, there exists
a matrix $\pm A \in SL(N, \Z)$ such that the vector fields $\hat \psi_i, 0<i\le N$
defined as in eq.~\eqref{hatpsi} satisfy $\hat \psi_i |_x = 0, r<i\le N$ and such that
$\hat \psi_i|_x, 0<i\le r$ span the tangent space $T_x O_x$. We start our discussion
with a general lemma.

\begin{lemma}\label{lemma2}
Let $\mL \subset \K=\T^{N}$ be an $n$-dimensional closed subgroup. Then there
are matrices of integers $(A_{ij})_{i,j=1}^N$ and $(v_{ij})_{i,j=1}^r$ where $r=N-n$ and $\det A = \pm 1$,
with the property that $\mL = \alpha_A(\mL_0 \times \mL_1)$. Here
\bena\label{groups}
\mL_0 &=& \{0_{r}\} \times \mr^{N-r}/\Lambda_{N-r} \, ,\\
%\Big\{
%\sum_{i=r+1}^N \tau_i \underline b_i \, {\rm mod} \, \Lambda_N \,\, \Big| \,\, \tau_i \in \mr
%\Big\}\\
\mL_1 &=& (v^{-1} \Lambda_r)/\Lambda_r \times \{ 0_{N-r} \} \, ,
%\Big\{
%\sum_{i,j=1}^r (v^{-1})_{ij} \tau_i \underline b_{j} \, {\rm mod} \, \Lambda_N \,\, \Big| \,\, \tau_i \in \Z
%\Big\} \, .
\eena
where $\Lambda_r$ has been identified with the lattice generated by $\underline b_1, \dots, \underline b_r$ with
origin denoted $0_r$, and where
$\Lambda_{N-r}$ has been identified with the lattice generated by $\underline b_{r+1}, \dots, \underline b_N$,
with origin denoted $0_{N-r}$. We have also written $v^{-1} \Lambda_r$ for the lattice of $\mr^r$ generated by
$\sum_{j=1}^r (v^{-1})_{ij} \underline b_j$, where $i=1, \dots, r$.
Hence $\mL_0$ is connected, $\mL_1$ is finite,
\ben
\mL_1 \cong \Z_{p_1^{\alpha_1}} \times \dots \times \Z_{p_M^{\alpha_M}}   \, , \quad |\mL_1| = p_1^{\alpha_1} \dots p_M^{\alpha_M} = |\det (v_{ij})_{i,j=1}^r| \, ,
\een
with $p_j>0$ prime.
\end{lemma}
{\em Proof:} Let us first assume that $\mL$ is also connected.
Then $\mL$ is a compact, abelian, connected Lie-group and so must be isomorphic
to $\T^n$. Let $\beta: \T^n \to \mL$ be the isomorphism. We identify
$\K = \T^N$ with $\mr^N/\Lambda_N$, where $\Lambda_N$ is the standard
lattice. Similarly, we identify $\T^n$ with
$\mr^n/\Lambda_n$, with $\Lambda_n = {\rm span}_\Z (\underline b_i)_{i=r+1}^N$.
Let $\underline a_i = \beta({\underline b}_i) \in \Lambda_N$,
where $i=r+1, \dots, N$. If $\lambda_i \in \mr$ are such that
\bena
\underline c &=& \lambda_1 \underline a_{r+1} + \dots + \lambda_n \underline a_N\nonumber\\
              &=& \beta(\lambda_1 \underline b_{r+1} + \dots + \lambda_n \underline b_N) \in \Lambda_N \, ,
\eena
then it follows that $\lambda_i \in \Z$. We conclude from \cite[Cor. 3, I.2.2]{cassels}
that there are vectors $\underline a_{1}, \dots, \underline a_{r} \in \Lambda_N$
such that $\underline a_{1}, \dots, \underline a_{N}$ form a basis of $\Lambda_N$.
We now let $A$ be the $N \times N$ matrix of integers such that $\underline b_i A^T = \underline a_i$ for
$i=1, \dots, N$.
Then $\det A = \pm 1$ because the matrix relates two bases of the lattice $\Lambda_N$. Since
$\mL_0$ viewed as a subgroup of $\T^N$ is generated precisely by $\underline b_{r+1}, \dots, \underline b_N$,
this proves the lemma when $\mL$ is connected.

In the general case, $\mL$ is isomorphic to the cartesian product of a torus
and cyclic groups of order given by a prime power, i.e. there is an isomorphism $\beta: \T^n \times \prod \Z_{p_j^{\alpha_j}} \to \mL$.
For $j=1, \dots, M$, let $\underline c_j$ be the image under $\beta$ of the
generator of the $j$-th cyclic finite group in the decomposition, projected onto the (real) span
of $\underline a_{1}, \dots, \underline a_r$. The vectors $\underline c_1, \dots, \underline c_M$
together with $\underline a_{1}, \dots, \underline a_r$ generate an $r$-dimensional lattice
$\Gamma_{r}$. Let $\underline \gamma_1, \dots, \underline \gamma_{r}$ be a basis of the lattice
$\Gamma_{r}$. It follows from \cite[Thm.~1,I.2.2]{cassels} that there are integers $v_{ij}$ such that
$v_{ii}>0, v_{ii} > v_{ji}$ for $j > i$, and
\bena\label{system}
\underline a_{1} &=& v_{11} \underline \gamma_1 \nonumber\\
\underline a_{2} &=& v_{21} \underline \gamma_1 + v_{22} \underline \gamma_2 \nonumber\\
\vdots && \vdots \nonumber\\
\underline a_r     &=& v_{r1} \underline \gamma_1 + v_{r2} \underline \gamma_2 + \dots + v_{rr} \underline \gamma_r \, .
\eena
It is evident that $\mL$ is given by the image under $\alpha_A$ of the cartesian product of the group given by the
real multiples of $\underline a_{r+1}, \dots, \underline a_N$ mod $\Lambda_N$
and the group of integer multiples of $\underline \gamma_1, \dots , \underline \gamma_r$ mod $\Lambda_N$.
The first group is the image under $\alpha_A$ of $\mL_0$, while the second is the image of $\mL_1$.
This proves that $\mL = \alpha_A(\mL_0 \times \mL_1)$.

From the system~\eqref{system} one sees that the order of $\mL_1$ is given by
$$
|\mL_1| = \prod_{i=1}^r v_{ii} = \det (v_{ij})_{i,j=1}^r \, .
$$
On the other hand, $\alpha_A^{-1} \circ \beta$ is an isomorphism between $\T^n \times \prod \Z_{p_j^{\alpha_j}}$
and $\mL_0 \times \mL_1$. The number of connected components of the first group is given by
$\prod p_j^{\alpha_j} = |\prod \Z_{p_j^{\alpha_j}}|$, while it is given by $|\mL_1|$ for the second. This finishes
the proof of the lemma. \qed.

\medskip
\noindent

We apply this lemma to the isotropy group $I_x \subset \K$, and we formulate the
 intermediate result as another lemma for future reference:

\begin{lemma}\label{lemma3}
Let $x \in S_r$. There are integer matrices $(v_{ij})_{i,j=1}^r$ and $(A_{ij})_{i,j=1}^N$
(depending on $x$)
with $\det A = \pm 1$ such that $I_x = \alpha_A(\mL_0 \times \mL_1)$, with $\mL_0$
and $\mL_1$ the groups given above in eq.~\eqref{groups}. Alternatively, we can say
that $I_x$ is generated by the elements
\ben
k(\tau_1, \dots, \tau_N) := \alpha_A \Big[ \frac{1}{2\pi} \Big( \sum_{i=r+1}^N \tau_i \underline b_i
+ \sum_{i,j=1}^r (v^{-1})_{ij} \tau_i \underline b_{j} \Big)
\Big] \, ,
\een
where $\tau_i \in \mr$ for $r+1 \le i \le N$, and where $\tau_i \in 2\pi \Z$ for
$1 \le i \le r$.
\end{lemma}
If we define $\hat \psi_i = \sum A_{ij} \psi_j$, then lemma~\ref{lemma3} implies that
$\hat \psi_i |_x = 0$ for $i=r+1, \dots, N$, and $\hat \psi_j |_x$ span
$T_x O_x$ for $j=1, \dots, r$.

We now continue our analysis by inspecting the action of $I_x$ on the tangent space $T_x \Sigma$.
Let $k \in I_x$. Then, because $k \cdot x = x$, this induces a linear map $k: T_x \Sigma \to T_x \Sigma$
satisfying $h(k \cdot X, k \cdot Y) = h(X,Y)$ for all $X,Y \in T_x \Sigma$. In fact, because
$k \cdot \psi_i = \psi_i$ for any of our Killing fields, it follows that $k$ leaves each vector in the tangent space
$T_x O_x$ invariant. But then it also leaves the orthogonal complement $W_x$ invariant. Let
$\{e_1, \dots, e_{s-r}\}$ be an orthogonal basis of $W_x$. So for every $k \in I_x$, we get a
representing orthogonal matrix $(k_{ij}), 0<i,j\le s-r$ acting on the orthognonal basis by
$k \cdot e_i = \sum k_{ij} e_j$. Because
$\Sigma$ is assumed to be orientable, we have a distinguished non-vanishing rank $s$
totally anti-symmetric tensor field $\epsilon$ (determined up to sign by $\epsilon_{a_1 \dots a_s} \epsilon_{b_1 \dots b_s}
h^{a_1 b_1} \dots h^{a_s b_s} = s!$). This tensor is invariant under the isometries of $\Sigma$, so in
particular $k \cdot \epsilon = \epsilon$ at point $x$, for any $k \in I_x$. Because $k \cdot \psi_i$ for any
of our Killing fields, this implies that the action of $k$ on $W_x$ preserves the orientation, so the matrix
$(k_{ij})$ representing this action has determinant ${\rm det} \, (k_{ij}) = +1$, and $(k_{ij}) \in SO(s-r)$. In particular,
$(k_{ij})$ must have an even number of $-1$ eigenvalues.
The matrices $(k_{ij})$ commute for different choices of $k \in I_x$, and so we may put them simultaneously into Jordon normal
form. By making a change of basis of
the $\{e_1, \dots, e_{s-r}\}$ with an orthogonal element $g \in O(s-r)$, we may achieve that
\ben\label{kR}
k \cdot (e_{2j-1} + ie_{2j}) = \e^{i\theta_j}(e_{2j-1} + ie_{2j}) \, , \quad 0<j\le [(s-r)/2] \quad
\text{if $s-r$ even}
\een
together with $k \cdot e_{s-r} = e_{s-r}$ when $s-r$ is odd\footnote{Here it has been used that
$(k_{ij})$ has determinant $+1$. Otherwise $(k_{ij})$ could also act as a reflection on an odd number of
basis vectors.}. The phases $\theta_j$ depend on $k$. For the elements of the isotropy
group given by lemma~\ref{lemma3}, we have in fact
\begin{multline}
k(0, \dots, 2\pi, \dots 0) \cdot (e_{2j-1} + ie_{2j}) \\= {\rm exp}
\Big(
2\pi i \sum_{m=1}^r (v^{-1})_{lm} w_{mj}
\Big)
(e_{2j-1} + ie_{2j}) \, , \quad 0<j\le [(s-r)/2]
\end{multline}
if $s-r$ is even
together with $k(0, \dots, 2\pi, \dots 0) \cdot e_{s-r} = e_{s-r}$ when $s-r$ is odd.
Here, the $2 \pi$ is in the $l$-th slot, with $l \le r$. The
$w_{ij}$ are integers, which follows from the fact that the
group elements $k(\sum_j v_{ij} \underline b_{j})$ are the identity, by lemma~\ref{lemma3}.
The above formula becomes somewhat more transparent if we note that the elements
$\underline \gamma_i = \sum_{j=1}^r (v^{-1})_{ij} \underline b_j$ defined for $i=1, \dots, r$
generate a copy of the isotropy subgroup $I_x \cong (v^{-1} \Lambda_r)/\Lambda_r
\cong \prod_j \Z_{p_j^{\alpha_j}} \cong \langle \underline \gamma_i \,\, {\rm mod} \, \Lambda_r \rangle$,
see lemma~\ref{lemma2}. Thus, we may
view the exponential expression in the above formula as a homomorphism
\ben\label{varthetadef}
\vartheta_j: (v^{-1} \Lambda_r)/\Lambda_r \to {\mathbb S}^1 = \{ z \in \mc \mid \,\, |z| = 1 \} \,\,\, ,
\quad \vartheta_j(\underline \gamma_k) = \e^{
2\pi i \sum_{m=1}^r (v^{-1})_{km} w_{mj}} \, .
\een
We also have
\ben
k(0, \dots, \tau_l, \dots 0) \cdot (e_{2j-1} + ie_{2j}) = {\rm exp}
(
i \tau_l w_{lj}
)
(e_{2j-1} + ie_{2j}) \, , \quad 0<j\le [(s-r)/2]
\een
together with $k(0, \dots, \tau_i, \dots 0) \cdot e_{s-r} = e_{s-r}$ when $s-r$ is odd. Here,
the $\tau_l$ is in the $l$-th slot, and $r+1 \le l \le N$. The $w_{ij}$ are again integers.

As yet, the basis $\{ e_1, \dots, e_{s-r} \}$ has only been defined in $W_x$, but we now wish to
define it for any $W_y$, with $y \in O_x$. Let
\ben\label{xtdef}
x(\tau_1, \dots, \tau_r) = k(\tau_1, \dots, \tau_r, 0, \dots, 0) \cdot x \, ,
\quad
0 \le \tau_i < 2\pi \, ,
\een
where $k(\underline \tau)$ is as in lemma~\ref{lemma3}.
Note that $x(\underline{\tau})$ is periodic in $\underline \tau$ with period $2\pi$ in each component of $\underline \tau$,
and that $\underline \tau \in [0, 2\pi)^r \to x(\underline{\tau}) \in O_x$ provide (periodic) coordinates in $O_x$.
We define our basis elements in $W_{x(\underline \tau)}$ by transporting $\{ e_1, \dots, e_{s-r} \}$
to $x(\underline \tau)$ with the group element in eq.~\eqref{xtdef}. We call this basis
$\{ e_1 (\underline \tau), \dots, e_{s-r} (\underline \tau)\}$. We note that this is still an orthonormal
system, because it was obtained by an isometry between $W_x \to W_{x(\underline\tau)}$.
Note that this basis is not periodic in $\underline \tau$, by eq.~\eqref{kR}. To obtain an orthonormal basis
$\{\tilde e_1(\underline \tau), \dots, \tilde e_{s-r}(\underline \tau)\}$ that is periodic in $\underline \tau$,
we set
\ben\label{tile}
\tilde e_{2j-1}(\underline \tau) + i
\tilde e_{2j}(\underline \tau) = {\rm exp}
\Big(
-i \sum_{m,l=1}^r \tau_l (v^{-1})_{lm} w_{mj}
\Big)
(e_{2j-1}(\underline \tau) + ie_{2j}(\underline \tau)) \, ,
\een
for $0<j\le [(s-r)/2]$, together with
$
\tilde e_{s-r}(\underline \tau) = e_{s-r}(\underline \tau)
$
when $s-r$ is odd.

In an open neighborhood of $O_x$, we now define coordinates as follows. First, {\em on} $O_x$, we
use the coordinates $(y_{s-r+1}, \dots, y_s) \in [0, 2\pi)^r \mapsto x(y_{s-r+1}, \dots, y_s)$. In a neighborhood of $O_x$
we use
\ben\label{cots}
(y_1, \dots, y_{s}) \mapsto {\rm Exp}_{x(y_{s-r+1}, \dots, y_s)}
\left(
\sum_{j=1}^{s-r} y_{j} \tilde e_j(y_{s-r+1}, \dots, y_s)
\right) \, .
\een
Here, ``Exp'' is the exponential map for our metric $h$, i.e., $(y_{1}, \dots, y_{s-r})$ are Riemannian
normal coordinates transverse to $O_x$. They cover an open neighborhood of $O_x$.
From the construction of the coordinates,
the action of the isometry group $\K$ in these coordinates is described by the following lemma:
\begin{lemma}\label{lemma4}
Let $x \in S_r$, let $(v_{ij})$ be the matrix and $k(\tau_1, \dots, \tau_N) \in I_x$ be
as in lemma~\ref{lemma3}.
Then, in terms of the coordinates~\eqref{cots} covering a neighborhood of
$O_x$, the action of $\K$ is given by
\bena\label{I}
&&k(\sigma_{1}, \dots, \sigma_{r}, 0, \dots, 0) \cdot
( y_1 + iy_2, \dots, y_{s-r-1} + iy_{s-r}, y_{s-r+1}, \dots, y_s)\\
&=&\Big(
(\exp\Big[
i\sum_{l,m=1}^r \sigma_{l} (v^{-1})_{lm} w_{mj} \Big](y_{2j-1} + iy_{2j}) )_{j=1}^{[(s-r)/2]}, \,\,
(y_{s-r+i} + \sigma_i)_{i=1}^r  \Big)\nonumber
\eena
when $s-r$ is even. When $s-r$ is odd, $y_{s-r}$ remains unchanged. Furthermore,
\bena\label{II}
&&k(0, \dots, 0, \sigma_{r+1}, \dots, \sigma_{N})
\cdot (y_1 + iy_2, \dots, y_{s-r-1} + iy_{s-r}, y_{s-r+1}, \dots, y_s)\\
&=&
\Big(
(\exp\Big[
i\sum_{l=r+1}^N \sigma_{l}  w_{lj} \Big](y_{2j-1} + iy_{2j}) )_{j=1}^{[(s-r)/2]}, \,\,
(y_{s-r+i})_{i=1}^r \Big) \nonumber
\eena
when $s-r$ is even. When $s-r$ is odd, $y_{s-r}$ remains unchanged.
\end{lemma}
Let $A$ be the matrix in lemma~\ref{lemma4}, and let $\hat \psi_i = \sum_j A_{ij} \psi_j$. By
lemma~\ref{lemma4}, the Killing fields $\hat \psi_i$ are related to the coordinate vector fields $\partial_{y_i}$ as:
\ben
\left(
\begin{matrix}
\hat \psi_1\\
%\hat \psi_2\\
\vdots\\
\hat \psi_r\\
\hat \psi_{r+1}\\
\vdots\\
\hat \psi_N
\end{matrix}
\right)
=
\left(
\begin{matrix}
v_{11} & \dots & v_{1r}   &   w_{1\, 1} & \dots & w_{1 \, [(s-r)/2]}\\
%0   & p_2 & \dots & 0 &   w_{2\, 1} & \dots & w_{2\, [(s-r)/2]}\\
\vdots &        & \vdots & \vdots &    & \vdots \\
v_{r1} &  \dots        & v_{rr}      & w_{r\, 1} & \dots & w_{r \, [(s-r)/2]}\\
0 &  \dots        & 0      & w_{r+1 \, 1} & \dots & w_{r+1 \, [(s-r)/2]}\\
\vdots &         & \vdots & \vdots &    & \vdots \\
0 &  \dots        & 0      & w_{N \, 1} & \dots & w_{N \, [(s-r)/2]}\\
\end{matrix}
\right)
\left(
\begin{matrix}
\partial_{y_{s-r+1}} \\
%\partial_{\tau_2} \\
\vdots\\
\partial_{y_s} \\
y_1 \, \partial_{y_2} - y_2 \, \partial_{y_1}\\
\vdots\\
y_{s-r-1} \, \partial_{y_{s-r}} - y_{s-r} \, \partial_{y_{s-r-1}}
\end{matrix}
\right)
\een
when $s-r$ is even. When $s-r$ is odd, there is an analogous expression.
Let us denote the $N \times (r+[(s-r)/2])$ matrix in this expression as $C$.  When $N-r = [(s-r)/2]$,
$C$ is a square $N \times N$ matrix. Furthermore, each
of the commuting, locally defined Killing fields $\partial/\partial y_i$ and $y_{2j-1} \partial/\partial y_{2j} -
y_{2j} \partial/\partial y_{2j-1}$ on the right side of the above equation is periodic, with period precisely
$2\pi$. Hence, when $N-r = [(s-r)/2]$, the matrix $C$ must have determinant $\pm 1$. So we get the condition
\ben
\det \, (v_{ij})_{i,j=1}^r  \cdot \det \, (w_{(r + i)j})_{i,j=1}^{N-r} = {\rm det} \, C = \pm 1.
\een
Because both determinants on the left are integers, we conclude we conclude that
they must be $\pm 1$. In view of lemma~\ref{lemma2}, this means
$p_1 = \dots = p_r = 1$ and ${\rm det} \, (w_{(r + i) \, j})_{i,j=1}^{N-r}=\pm 1$. We summarize our findings in
another lemma:
\begin{lemma}\label{lemma5}
Let $\psi_1, \dots, \psi_N$ be Killing fields as above, $x \in S_r$, $n=N-r = [(s-r)/2]$. Then
$p_1 = \dots = p_r = 1$ (see lemma~\ref{lemma2}), and
${\rm det} \, (w_{(r + i) \, j})_{i,j=1}^{n} = \pm 1$. Furthermore, in that
case $I_x$ is connected.
\end{lemma}

With the help of the above lemmas, we are now ready to analyze the orbit space $\hat \Sigma$ in the
case when $N=s-2$. We first cover
$\Sigma$ by the coordinate systems defined in eq.~\eqref{cots}. Within each such coordinate system,
we can then separately perform the quotient by $\K$. We need to distinguish the cases $n=0,1,2$, where
$n=s-2-r$, and where the coordinate system covers a point $x \in S_r$.

\medskip
\noindent
{\sl Case 0:} For $n=0$ and hence $r= s-2$, the isotropy group $I_x$ is discrete and is isomorphic to
the group generated by the elements $\underline \gamma_i = \sum_{j=1}^{s-2} (v^{-1})_{ij} \underline b_j$,
see lemmas~\ref{lemma2},~\ref{lemma3}. It is also isomorphic to $\prod_j \Z_{p_j^{\alpha_j}}$.
Furthermore, by combining lemmas~\ref{lemma3} and~\ref{lemma4}, the action of these isotropy group
in a neighborhood of $O_x$ can be written
as
\ben
k(0, \dots, 2\pi, \dots, 0) \cdot ( y_1 + iy_2,y_3, \dots, y_s)\\
= \Big(
\vartheta(\underline \gamma_j) (y_1 + iy_2), y_3, \dots, y_s\Big) \, ,
\een
where we are using the notation introduced in eq.~\eqref{varthetadef} for the homomorphism
$\vartheta: \prod_j \Z_{p_j^{\alpha_j}} \to {\mathbb S}^1$, and where the ``$2\pi$'' is in the $j$-th slot.
Consider now the kernel ${\rm ker} \, \vartheta$.
If $g$ is an element in its kernel, then it is evident from the above formula
that the corresponding isometry of $\Sigma$ acts by
the identity both in a full neighborhood of $O_x$. Consequently, $g$ must be the identity element of
the group, since we are assuming the action to be effective. In particular, $\vartheta$ is injective.
Consider next the image ${\rm ran} \, \vartheta$. This is a finite subgroup of the circle group ${\mathbb S}^1$.
Hence it is given by ${\rm ran} \, \vartheta = \{ \e^{2\pi i k/q} \mid k=0, \dots, q-1\} \cong \Z_q$
for some $q$. It follows from the fact that $\vartheta$ is injective that
\ben
|{\rm ran} \, \vartheta| = q = |\prod_j \Z_{p_j^{\alpha_j}}| = \prod_{j} p^{\alpha_j}_j \, .
\een
Furthermore, it follows that the inverse $\vartheta^{-1}$ is a well-defined map on $\Z_q$,
which can be viewed as taking values in the isotropy group $I_x \subset \K$.

It follows from the discussion that, within the neighborhood considered, the quotient is modeled upon
$\mr^2/\Z_q$, where $q=\prod p_j^{a_j} = | \det v\, |$ (see lemma~\ref{lemma2},~\ref{lemma3}),
and where the cyclic group of $q$ elements acts on the
coordinates $y_1+iy_2$ by complex phases $\e^{2\pi i/q}$. Thus, in a neighborhood of $O_x$, the quotient
space is an orbifold $\mr^2/\Z_q$. In particular, we see that the orbits having non-trivial discrete
isotropy group must be isolated points in $\hat \Sigma$. These orbits are also called ``exceptional orbits''.
The other orbits in case (0) have no isotropy group and are called ``principal orbits''.

\medskip
\noindent
{\sl Case 1:} For $n=1$, lemma~\ref{lemma5} applies and $p_i=1$ for all $i$. We first factor by the group elements $[0, \dots, 0, \sigma_{s-2}]$, see
eq.~\eqref{II}, and afterwards by the group elements $[\sigma_1, \dots, \sigma_{s-3}, 0]$, see
eq.~\eqref{I}.
Then it is quite clear that the resulting  quotient space of our neighborhood
of $O_x$ is locally modeled upon $\mr \times \mr_{>0}$.
The first factor corresponds to the variable $y_3$, while the second factor to the variable
$\sqrt{y_1^2+y_2^2}$.

\medskip
\noindent
{\sl Case 2:} For $n=2$, lemma~\ref{lemma5} applies and $p_i=1$ for all $i$.
We first factor by the group elements $[0, \dots, 0, \sigma_{s-3}, \sigma_{s-2}]$, see
eq.~\eqref{II}, and afterwards by the group elements $[\sigma_1, \dots, \sigma_{s-4}, 0,0]$, see eq.~\eqref{I}.
Then it is quite clear that the resulting quotient space of our
neighborhood of $O_x$ is locally modeled upon $\mr_{>0} \times \mr_{>0}$.
The first factor corresponds to the variable $\sqrt{y_3^2+y_4^2}$, while the second factor to the variable
$\sqrt{y_1^2+y_2^2}$.

Thus, we have proven the following theorem:
\begin{thm}\label{thm1}
Let $\Sigma$ be an orientable connected $s$-dimensional Riemannian manifold with $s-2$ pairwise commuting
Killing fields generating an action of the group $\K = \T^{s-2}$ by isometries.
Then the quotient space $\hat \Sigma = \Sigma/\K$ is an orbifold with conical singularities, boundary segments, and corners.
Thus, each point of $\hat \Sigma$ has a neighborhood modeled on a neighborhood of the tip of
a cone $\mr^2/\Z_q$, on a half-space $\mr \times \mr_{>0}$, or on a corner,
$\mr_{>0} \times \mr_{>0}$. In the first case, the corresponding isotropy group is finite
and $q$ is given by the order of this group.
\end{thm}

Each point of the boundary segments, corners, or orbifold points in $\Sigma$ is associated with
an isotropy group $I_x$ as in lemma~\ref{lemma3}. It follows from our discussion in case 1) that, as
long as we stay within one boundary segment, the isotropy group does not change. Furthermore, by
lemmas~\ref{lemma5} and~\ref{lemma3}, the isotropy group $I_x$ is connected for points $x$
associated with boundaries and corners. For $x$ associated with conical singularities, $I_x$ is discrete,
again by lemmas~\ref{lemma5} and~\ref{lemma3}.
It also follows from our discussion of cases 1) and 2) that, for each boundary segment and each corner, the isotropy group is  completely characterized by
an integer matrix $A$ of determinant $\pm 1$. Furthermore, it follows from our discussion in case 0) that the isotropy
group $I_x$ is characterized by an integer $q$ and an injective homomophism $\vartheta^{-1}: \Z_q \to \T^{s-2}$,
whose image is $I_x$.
There is one such matrix $A$ for each boundary
segment one for each corner, and one such $q, \vartheta^{-1}$ for each conical singularity.
The matrices $A$ are actually not completely characterized by
the corresponding isotropy subgroup $I_x$. In fact, by lemma~\ref{lemma2} (with $\mL=I_x$, $x \in S_r$) the position of
the isotropy subgroup within $\K$ is uniquely determined by the class ($N = s-2$)
\ben
[A] \in \frac{SL(N, \Z)}{U(N-r, r; \Z)}
\een
where $U(N-r,r; \Z)$ is the group of block-upper triangular matrices with block sizes $N-r,r$ with
integer entries and determinant $\pm 1$. The quotient by such matrices $U$ takes
into account the fact that left-multiplying an $A$ by such a matrix
gives the same isotropy subgroup. When $N-r = n = 1$ (corresponding to case 1, and a boundary
segment), the class of $A$ is determined by the last row $(a_{N1}, ..., a_{NN})$ of the matrix $A$, and
we have $\sum a_{Ni} \psi_i |_x = 0$ for each point $x$ in $M$ corresponding to the boundary segment under
consideration.
When $N-r = n = 2$ (corresponding to case 2, and a corner), the class of $A$ is determined
by the last two rows $( a_{(N-1)1},  \dots,  a_{(N-1)N}), (
a_{1N},  \dots,  a_{NN})$ up to a $SL(2, \Z)$ transformation acting on each column of the $N \times 2$ matrix formed
from these.
We have $\sum a_{(N-1)i} \psi_i |_x = 0$ and $\sum a_{Ni} \psi_i |_x = 0$ for each point $x$ in $\Sigma$ corresponding to the
corner under consideration.

\medskip

If $\{I_j\} \subset \partial \hat \Sigma$ is the collection of boundary segments, and if $I_{ij} = I_i \cap I_j$ are
the corresponding corners, then for each $I_i$, we have a vector $\underline a(I_i) \in \Z^N$ which is
the last row of the matrix $A$ corresponding to that boundary segment. The greatest common divisor (g.c.d.) of
the entries of the vector may be assumed to be equal to 1,
\ben\label{normalization}
{\rm g.c.d.}\{ a_i(I_j) \mid i=1, \dots, D-3 \} = 1 \, ,
\een
For each corner $I_{ij}$, the corresponding vectors $\underline a(I_i)$ and $\underline a(I_{j})$ must be such that
the $N \times 2$ matrix formed from these vectors can be supplemented by $N-2$ rows of integers
to an $SL(N, \Z)$-matrix, and this introduces a constraint on the pair $\underline a(I_i), \underline a(I_j)$.
In the case $s=4$ (i.e., $N=2$), the constraint at each corner $I_{ij}$ is simply that $\det \, (\underline a(I_i), \underline a(I_j)) = \pm 1$. In
the general, the constraint on the vectors adjacent to a corner $I_{ij}$ can be restated as
follows applying~\cite[Lemma 2, I.2.3]{cassels}:

\begin{proposition}\label{proposition1}
Let $\{I_j\} \subset \partial \hat \Sigma$ be the boundary
segments. With each boundary segment there is associated
a vector $\underline a(I_j) \in \Z^{s-2}$ and
$\sum a_i(I_j) \psi_i=0$ at the corresponding points of $\Sigma$.
At a corner $I_{ij} = I_i \cap I_j$, the vectors are subject to the
constraint
\ben\label{consistency}
{\rm g.c.d.} \left\{ Q_{kl} \mid 1 \le k < l \le D-3 \right\} = 1 \, .
\een
Here, the numbers $Q_{kl} \in \Z$
are defined by
\ben\label{consistency1}
Q_{kl} = |\det \left(
\begin{matrix}
a_k(I_i) & a_k(I_j)\\
a_l(I_i) & a_l(I_j)
\end{matrix}
\right)| \, .
\een
Let $\{ \hat x_i \} \subset \hat \Sigma$ be the conical singularities.
With each one, there is associated a natural number $q_i > 1$, specifying
the type $\mr^2/\Z_{q_i}$ of the conical singularity, and a homomorphism
$\vartheta^{-1}_i: \Z_{q_i} \to \K$, whose image is the discrete isotropy
subgroup at $x_i=$ any point in $\Sigma$ in the class $\hat x_i \in \hat \Sigma$.
\end{proposition}

\paragraph{Remarks:} (1) The data consisting of (i) the vectors $\{\underline a(I_j)\}$,
(ii) the pairs $\{q_j, \vartheta^{-1}{}_j\}$, (iii) the orientation of $\hat \Sigma$,
(iv) the topological type of $\hat \Sigma$ (genus) has been called the
``weighted orbit space'' by Orlik and Raymond~\cite{orlik1, orlik2} for the case $s=4$.
Our proposition hence may be viewed as a generalization of their analysis
to higher dimensions. \\
\noindent
(2) If the boundary $\partial \hat \Sigma$ is empty, then, as explained
in detail in~\cite[Sec.~1.3]{orlik1}, there are additional invariants associated
with the $\K$-space $\Sigma$. These may be characterized as obstructions to lift
certain cross sections on the boundaries of tubular neighborhoods of the
orbifold-type orbits $\hat x_i$ to $\Sigma$ and may be thought of as a class in the
space
\ben
H^2 \left(\hat \Sigma, \bigcup_{i=1}^m D^2_i ; \Z^{s-2} \right) \cong \Z^{s-2} \,
\een
where each $D_i^2$ is a disk around $\hat x_i$. This class has to be added to the data.

\medskip
\noindent

By a similar analysis we can also prove the following theorem on cohomogeneity-1 torus actions:

\begin{thm}\label{thm2}
Let $(\H, \gamma)$ be a connected, orientable, compact Riemannian manifold
of dimension $s-1>1$ with an isometry group containing an $(s-2)$-dimensional
torus $\K=\T^{s-2}$. Then the orbit space $\hat \H = \H/\K$ is diffeomorphic to
a closed interval as a manifold with boundary, or to a circle. In the first case, we have the following
possibilities concerning the topology of $\H$:
\ben\label{htopology}
\H \cong
\begin{cases}
\S^2 \times \T^{s-3}\\
\S^3 \times \T^{s-4}\\
L(p,q) \times \T^{s-4}
\end{cases}
\een
Here $L(p,q)$ is a 3-dimensional Lens space. In the second case, $\H \cong \T^{s-1}$.
\end{thm}

{\em Proof}:
Let $\psi_i, i=1, \dots, s-2$ be the commuting Killing fields of period $2\pi$ generating
the action of $\K$ on $\H$.
In the decomposition $\H = \cup S_r$ defined as in eqs.~\eqref{srdef}, \eqref{decompdef},
only the sets with $r=s-1$ and $r=s-2$ may be non-zero, by lemma~\ref{lemma1}. We consider
these cases separately.

{\sl Case 0):} Let $x \in S_{s-1}$, and let $T_x \H = T_x O_x \oplus W_x$
be the orthogonal decomposition into vectors tangent to $O_x$ and those orthogonal to $O_x$.
By assumption, the dimension of $W_x$ is one. If $k \in I_x$ is in the isotropy group, then
it leaves $T_x O_x$ invariant, as $k \cdot \psi_i = \psi_i$ for all $i$. So $k$ acts as $\pm 1$ on
$W_x$. But $k$ also preserves the rank $(s-1)$ anti-symmetric tensor $\epsilon$ compatible
with the metric, which exists since $\H$ is orientable. So $k$ acts as $+1$ on $W_x$, and
hence as the identity on $T_x \H$. The action of $k$ must hence leave invariant any piecewise
smooth geodesic on $(\H, \gamma)$ through $x$, and therefore $k$ must act as the identity on
all of $\H$, since this is a connected manifold. Thus, the isotropy group $I_x$ is trivial
in case 0). Consequently, near $O_x$, $\hat \H = \H/\K$ has the structure of a 1-dimensional
manifold, i.e., an open interval.

{\sl Case 1):} Let $x \in S_{s-2}$. By exactly the same arguments as given above using
lemmas~\ref{lemma4} and~\ref{lemma5}, the action of $\K$ is given near $O_x$ in local
coordinates $(y_1, \dots, y_{s-1})$ by
\bena
&&k(\sigma_1, \dots, \sigma_{s-2}) \cdot (y_1 + iy_2, y_3, \dots, y_{s-1}) \nonumber\\
&=& \Big(\exp\Big[
i\sum_{l=1}^{s-2} w_l \sigma_l \Big] (y_1 + iy_2), y_3 + \sigma_1, \dots, y_{s-1} + \sigma_{s-3} \Big) \, .
\eena
Here, $\pm A$ is some $SL(s-2, \Z)$ matrix, the numbers $w_l$ are integers, and $w_{s-2}=\pm 1$ (see lemma~\ref{lemma5}).
It is evident from this that $\sqrt{y_1^2+y_2^2}$ furnishes a coordinate for
$\hat \H$ in a neighborhood of $O_x$, thus identifying this neighborhood locally with
a half-open interval.

Because $\hat \H$ can be covered by neighborhoods of the kind described in cases 0)
and 1), i.e., open and half open intervals, and because $\hat \H$ is compact in a
natural topology and connected, it follows that $\hat \H$ must be a 1-dimensional
connected compact manifold with or without boundaries. In the first case, $\hat \H$ is
diffeomorphic to a closed interval, in the second case to a circle. In the first case, the two boundary points of this closed interval correspond to orbits $O_x$ respectively $O_y$ in $\H$
where an integer linear combination $\sum a_{i,1} \psi_i$ respectively $\sum a_{i,2} \psi_i$
vanishes.
%We can argue as in the proof of lemma~\ref{lemma2} and find a matrix
%$B \in SL(s-2, \Z)$ with the property that $\underline a B^T= (1, 0, \dots, 0)$ and
%$\underline b B^T= (p,q, 0, \dots, 0)$.
We can redefine our action of $\K$
using instead the Killing fields $\hat \psi_i = \sum A_{ij} \psi_j$ for some integer
matrix $A$ with $\det \, A = \pm 1$ in such a way that
on $O_x$ we have $\hat \psi_1 = 0$, while
on $O_y$ we have $p \hat \psi_1 + q \hat \psi_2 = 0$. Consider now the
subgroup $\mL \subset \K$ generated by $\hat \psi_3, \dots, \hat \psi_{s-2}$. Clearly,
$\mL$ is isomorphic to $\T^{s-4}$. It follows from the discussion of the cases 0) and 1)
that there are no points in $\H$ which are fixed under a non-trivial element of $\mL$, so
$\H \cong (\H/\mL) \times \T^{s-4}$. Then, $\H/\mL$ is a three-dimensional manifold on
which there acts the subgroup of isometries in $\K$ generated by $\hat \psi_1, \hat \psi_2$.
It is not difficult to see, and argued carefully in~\cite{Hollands:2007aj}, that
$\H/\mL$ is isomorphic to $\S^3$ if $(p,q) = (0,1)$, isomorphic to $\S^2 \times \T^1$ if
$(p,q) = (1, 0)$, and a Lens-space $L(p,q)$ otherwise.

In the second case, $\H$ must be diffeomorphic to the direct product of $\K$ and a circle, i.e.
to $\T^{s-1}$.
\qed

\subsection{The fundamental group of $\Sigma$}

In the previous section, we have analyzed oriented $s$-dimensional manifolds $\Sigma$ with an effective
action of $\K = \T^{s-2}$. We showed that the quotient space $\hat \Sigma = \Sigma/\K$
was an orientable 2-manifold with a finite number of conical singularities in the interior, and with
boundaries and corners. With each of the conical singularities $\hat x_i \in \hat \Sigma$ there
was associated an integer $q_i \in \Z$ and an injective homomorphism $\vartheta_i^{-1}: \Z_{q_i} \to \K$.
These homomorphisms may be written as
\ben\label{homoj}
\vartheta^{-1}_j(\e^{2\pi i/q_j}) = (\e^{2\pi i p_{1,j}/q_j}, \dots, \e^{2\pi i p_{s-2,j}/q_j}) \, ,
\een
where ${\rm g.c.d.}\{q_j, {\rm g.c.d.}\{p_{1, j}, \dots, p_{s-2,j} \}\} = 1$.
Furthermore, with each of the boundary intervals $I_i \subset \partial \Sigma$, there was
associated a vector $\underline a_i = (a_{1,i}, \dots, a_{s-2,i}) \in \Z^{s-2}$. On a corner, the vectors are subject
to the constraint~\eqref{consistency},~\eqref{consistency1}. If $\Sigma$ is compact, then $\hat \Sigma$ is
a compact oriented 2-dimensional topological manifold, and hence topologically of the form
\ben\label{hatsigma1}
\hat \Sigma \cong \hat \Sigma_g \setminus \bigcup_{j=1}^d D^2_j
\een
where each $D^2_j$ is a 2-dimensional disk, and where $\hat \Sigma_g$ is a closed Riemann surface of genus $g$.

One can show that the manifold $\Sigma$ with $\K$-action is fixed up to equivariant isomorphism by the data consisting of
$\hat \Sigma, \{I_i\}, \{\hat x_i\}, \{q_i, \underline p_i\}, \{\underline a_i\}$; we will indicate how
to prove this in subsection~\ref{subsec34}. Therefore, any topological invariant of $\Sigma$ must be
expressible in terms of these data.  It is evident that
the fundamental group $\pi_1(\Sigma)$ should provide a strong invariant for the topology of $\Sigma$.
It is given in the next theorem:

\begin{thm}\label{fundathm}
Let $\Sigma$ be a compact orientable manifold with an effective action of $\K = \T^{s-2}$ such that
$\partial \hat \Sigma \neq \emptyset$. Then the fundamental group can be presented as:
\bena\label{pione}
\pi_1(\Sigma) &=& \Big\{ k_1, \dots, k_{s-2}, d_1, \dots, d_c, h_1, \dots, h_d, m_1, \dots, m_g, l_1, \dots, l_g \Big| \nonumber\\
&& [m_1, l_1] \cdots [m_g,l_g] \cdot d_1 \cdots d_c \cdot h_1 \cdots h_d \, ; \nonumber\\
&& [m_i, k_j] \, ; \,  [l_i, k_j] \, ; \, [d_i, k_j] \, ; \, [h_i, k_j] \, ; \, [k_i, k_j] \, ;  \nonumber \\
&& k_1^{a_{1,1}} \cdots k_{s-2}^{a_{s-2,1}} \, , \dots, \,
k_1^{a_{1,b}} \cdots k_{s-2}^{a_{s-2,b}} \, ;
\nonumber\\
&& d^{q_1}_{1} k_1^{p_{1,1}} \dots k_{s-2}^{p_{s-2,1}} \, , \dots, \,
d^{q_c}_{c} k_1^{p_{1,c}} \dots k_{s-2}^{p_{s-2,c}}
\Big\} \, .
\eena
Here, we are using the usual notation for a finitely generated group in terms of its relations,
and $[x,y] = xyx^{-1}y^{-1}$ is the commutator of group elements.
Above, $g$ is the number of handles of $\hat \Sigma$, $c$ is the
number of conical singularities, $b$ is the number of intervals $\{I_i\}$ , and $d$ is the number
of boundary components in $\partial \hat \Sigma$ homeomorphic to circles, see eq.~\eqref{hatsigma1}.
\end{thm}

\noindent
{\em Proof:} The proof is essentially an application of the Seifert-Van~Kampen theorem, which
is described e.g. in~\cite[Chap.~4]{Massey}.
Let $x \in \Sigma$ be any point with trivial isotropy group, and let $k_i, i=1, \dots, s-2$ be the closed loops obtained
by a applying the $i$-th generator of $\pi_1(\K)$ (=generator of the $i$-th copy of $\T^1$ in
$\T^{s-2}$) to $x$. Let $d_i, i=1, \dots, c$ be lifts of loops going around the $i$-th conical singularity in $\partial \hat \Sigma$, and let $h_i, i=1, \dots, d$ be lifts of loops going around the $i$-th hole of $\hat \Sigma$ (=boundary component in $\partial \hat \Sigma$). We cut out a small disk $D^2_i$ around each of the conical singularities in $\hat \Sigma$,  we cut out a
small neighborhood of the boundary in $\hat \Sigma$, and we consider the corresponding subset of $\Sigma$. This
subset will have a homotopy group generated by $k_1, \dots, k_{s-2}, d_1, \dots, d_c, h_1, \dots, h_d$,
and generators $m_1, l_1, \dots, m_g, l_g$ corresponding to the $g$ handles of $\hat \Sigma$. The relations are
\ben
[m_1, l_1] \cdots [m_g,l_g] \cdot d_1 \cdots d_c \cdot h_1 \cdots h_d \, ; \,
[m_i, k_j] \, ; \,  [l_i, k_j] \, ; \, [d_i, k_j] \, ; \, [h_i, k_j] \, ; \, [k_i, k_j] \, .
\een
We now glue back in the neighborhood of the boundary. Since, near the $i$-th boundary segment $I_i$,
the generator $k_1^{a_{1,i}} \cdots k_{s-2}^{a_{s-2,i}}$ shrinks to zero size, we receive the relations
\ben
k_1^{a_{1,1}} \cdots k_{s-2}^{a_{s-2,1}} \, ; \, \dots \, ; \,
k_1^{a_{1,b}} \cdots k_{s-2}^{a_{s-2,b}} \,
\een
via the Van~Kampen theorem.
We finally glue in the disks around the conical singularities, each of which corresponds to a tube
$D^2 \times \T^{s-2}$. We must perform the gluing in such a way that the standard action of $\K$ on
$D^2 \times \T^{s-2}$ matches up with the action of $\K$ on $\Sigma$ near the exceptional orbits.
This action is characterized by the homomorphism~\eqref{homoj} for the $j$-th tube; we receive the
relations
\ben
d^{q_1}_{1} k_1^{p_{1,1}} \dots k_{s-2}^{p_{s-2,1}} \, ; \, \dots \, ; \,
d^{q_c}_{c} k_1^{p_{1,c}} \dots k_{s-2}^{p_{s-2,c}}
\een
from this operation, again via the Van~Kampen theorem.
\qed

The theorem has an interesting corollary in $s=4$ if the action of $\K$ has a fixed point, i.e. when
the orbit space has a corner. The vectors associated with the intervals $I_i, I_{i+1}$ adjacent
to the corner, $\underline a_i, \underline a_{i+1}$, must then satisfy $\det (\underline a_i, \underline a_{i+1}) =
\pm 1$ [see eq.~\eqref{consistency}]. This imposes the relation $k_1 = k_2 = e$ in eq.~\eqref{pione}.
Then, if $\pi_1(\Sigma) = 0$, this will imply that $g=d=0$, and $q_1, \dots, q_c = 0$. In other words,
if $s=4$, if the action has fixed point, and if $\Sigma$ is simply connected, then there are
no conical singularities, i.e., exceptional orbits. This was first proved using methods from singular cohomology
in~\cite{orlik1}.

The above theorem has another related corollary which will be relevant below in our application to the structure of
black holes. Let $D^2 \subset \hat \Sigma$ be any disk in the interior of the orbit manifold not
intersecting any of the boundaries or conical singularities. Thus, the orbits are all $(s-2)$-dimensional
tori, with no fixed points. The inverse image of $D^2$ in $\Sigma$ is homeomorphic to $D^2 \times \T^{s-2}$,
with $\K$ acting on the second factor. Let us denote the generators of
$\pi_1(D^2 \times \T^{s-2})$ by $k_1, \dots, k_{s-2}$, which are the $s-2$ generators of $\pi_1(\T^{s-2}) = \Z^{s-2}$.
Without loss of generality, we may assume that $k_j$ are the image of the paths generated by
the action of the $j$-th copy on $\K = \T^{s-2}$ on a point $x \in D^2 \times \T^{s-2}$.

From the inclusion $f: D^2 \times \T^{s-2} \to \Sigma$, we get a corresponding homomorphism
$f_*: \pi_1(D^2 \times \T^{s-2}) \to \pi_1(\Sigma)$. The way we have set things up, we may assume that
$f_*(k_j) = k_j$, using the same notation and assumptions as in the above theorem~\ref{fundathm}.

\begin{lemma}\label{lemmapione}
If $f_* : \pi_1(D^2 \times \T^{s-2}) \to \pi_1(\Sigma)$ is surjective, then we have $g = d = 0, q_1 = \dots = q_c = 1$.
In other words, $\hat \Sigma$ is a topologically a disk, and there are no conical singularities.
\end{lemma}

\noindent
{\em Proof:}  Using eq.~\eqref{pione} and the formula $f_*(k_j) = k_j$, we see that
$f_* \pi_1(D^2 \times \T^{s-2})$ is a normal subgroup of $\pi_1(\Sigma)$.
By assumption, the factor group
$\pi_1(\Sigma)/f_* \pi_1(D^2 \times \T^{s-2})$ is trivial. From the quotient,
the group $\pi_1(\Sigma)$ [see eq.~\eqref{pione}] receives the additional relations $k_j = e$ for $j=1, \dots, s-2$.
This means that the factor group is isomorphic to
\bena
\pi_1(\Sigma)/f_* \pi_1(D^2 \times \T^{s-2})
&\cong& \Big\{ d_1, \dots, d_c, h_1, \dots, h_d, m_1, \dots, m_g, l_1, \dots, l_g \Big| \\
&& [m_1, l_1] \cdots [m_g,l_g] \cdot d_1 \cdots d_c \cdot h_1 \cdots h_d \, ; \,
%&& [m_i, k_j] \, ; \,  [l_i, k_j] \, ; \, [d_i, k_j] \, ; \, [h_i, k_j] \, ; \, [k_i, k_j] \, ;  \nonumber \\
%&& k_1^{a_{1,1}} \cdots k_{s-2}^{a_{s-2,1}} \, , \dots, \,
%k_1^{a_{1,b}} \cdots k_{s-2}^{a_{s-2,b}} \, ;
%\nonumber\\
%&&
d^{q_1}_{1} \, ; \, \dots \, ; \,
%k_1^{p_{1,1}} \dots k_{s-2}^{p_{s-2,1}} \, , \dots, \,
d^{q_c}_{c}
%k_1^{p_{1,c}}
%\dots k_{s-2}^{p_{s-2,c}}
\Big\} \, . \nonumber
%\Z^d \times \Z^{2g} \times \Z_{q_1} \times \dots \times \Z_{q_m} \, ,
\eena
Evidently, this group is non-trivial unless $g = d = 0, q_1 = \dots = q_c = 1$, from which the lemma follows.
\qed

\subsection{The orbit space of the domain of outer communication}
We next want to determine the orbit space of a $D$-dimensional asymptotically Kaluza-Klein stationary
black hole spacetime $(M,g)$ with $D-3$ axial Killing fields $\psi_i, i=1, \dots, D-3$
generating an (effective) action of $\K = \T^{D-3}$. Thus, the total group isometries is $\G = \K \times \mr$,
with $\mr$ the additive group generated by the asymptotic timelike Killing field $t$. We have
the following theorem:

\begin{thm}\label{thm3}
Let $(M,g)$ be a stationary, asymptotically Kaluza-Klein, $D$-dimensional vacuum black hole
spacetime with isometry group $\G=\mr \times \K$, satisfying the technical
assumptions stated in sec.~\ref{sec1}. Then the orbit space
$\hat M = \langle\!\langle M \rangle\!\rangle/\G$ of the domain of outer communication
is a 2-dimensional manifold with boundaries and corners homeomorphic to a half-plane.
In particular, there are no conical singularities in $\hat M$.
The possibilities for the horizon topology are eqs.~\eqref{htopology}, with $s=D-1$.
One of the boundary segments $I_j \subset \partial \hat M$ is the quotient of the horizon
$\hat H=H / \G$, while the remaining $I_j$ correspond
to the various ``axis'', where $\sum a_i(I_j) \psi_i = 0$. The vectors
$\underline a(I_j) \in \Z^{D-3}$ are subject to the constraint~\eqref{consistency}
on each corner $I_i \cap I_j$.
\end{thm}

\medskip
\noindent
{\bf Remark:} In the statement concerning the horizon topology, eq.~\eqref{htopology},
we do {\em not} mean that the torus factors (such as in $\H \cong \S^2 \times \T^{D-4}$) correspond
to the rotations in the extra dimensions
near infinity.

\medskip
\noindent
{\em Proof}: The ``structure theorem''~4.3 of~\cite{chrcos} states that
$\langle \! \langle M \rangle \! \rangle$ contains a smooth, spacelike,
acausal slice $\Sigma$ whose boundary is a cross section $\H$ of the horizon,
which is asymptotic to a $\tau = const.$ slice
in the exterior under the identification of the exterior with
(part of) $\mr^{s,1} \times \T^{D-s-1}$, which invariant under the action of $\K = \T^{D-3}$
and which is transversal to the orbits of $t$ represented by the factor $\mr$ in $\G$.
Furthermore, if $F^\tau$ is the flow of $t$, then $\langle \! \langle M \rangle \! \rangle =
\cup_\tau F^\tau(\Sigma)$. This result will allow us to reduce the proof of
thm.~\ref{thm3} to thm.~\ref{thm1}.

We first factor $\langle \! \langle M \rangle \! \rangle$ by $\mr$. We can identify the
resulting space with $(\Sigma, h)$, with $h$ the Riemannian metric induced from $g$.
This metric is asymptotic to the standard flat metric on $\mr^s \times \T^{D-s-1}$ ($s=1,2,3$ or $4$)
in the exterior region. Evidently, $\K$ acts as a group of isometries on $(\Sigma, h)$,
and $\Sigma$ contains no points with discrete isotropy group. For definiteness, we focus
on the case $s=4$, the other cases are similar. The action of $\K$ is then conjugate in the exterior region
to the standard action which
acts on $\T^{D-5}$ by rotations along the generators, and which acts on $\mr^4$ by rotations in
the $12$- and $34$-plane.

We divide $\Sigma$ up into two pieces $\Sigma_0 \cup \Sigma_\infty$. The region $\Sigma_\infty$ is
the asymptotic region, and $\Sigma_0$ is the rest. The split can be arranged so that both pieces are
separately invariant under the action of $\K$. $\Sigma_0$ is a compact manifold with boundary $\partial
\Sigma_0$ consisting of $\H$ and of a second boundary component $\cong \S^3 \times \T^{D-5}$ bordering
on $\Sigma_\infty$. The quotient of $\Sigma$ is given by the union of the quotients
$\hat \Sigma_0 = \Sigma_0/\K$ and $\hat \Sigma_\infty = \Sigma_\infty/\K$. The action of
$\K$ on the exterior region is conjugate to the action of $\K$ on $(\mr^4 \setminus \{
(x_1, x_2, x_3, x_4) \mid R < r \}) \times \T^{D-5}$, where $R$ is the standard radius on $\mr^4$.
So the quotient is given by $\hat \Sigma_\infty \cong \{(R_1, R_2) \in \mr^2 \mid R_1, R_2 > 0, \,\, R_1^2 + R^2_2 > r^2\}$,
where $R_1$ can be identified with $\sqrt{x_1^2 + x_2^2}$ and $R_2$ with $\sqrt{x_3^2 + x_4^2}$. The boundary components of $\hat \Sigma_\infty$
defined by $R_i = 0, i=1,2$ correspond to an axis in the spacetime where $\psi_i, i=1,2$ vanish.
The quotient $\hat \Sigma_0$ can be determined as in thm.~\ref{thm1}, but we must now consider
a compact manifold $(\Sigma_0, h)$ with boundaries.
Near the boundary component $\cong \S^3 \times \T^{D-5}$ of $\partial\Sigma_0$, the quotient space
$\hat \Sigma_0$ must look like $\cong \{(R_1, R_2) \in \mr^2 \mid R_1, R_2 > 0, \,\, R_1^2 + R^2_2 \le r^2\}$.
Near the horizon boundary component $\H$, we can analyze the quotient space by combining the arguments in
thms.~\ref{thm1} and~\ref{thm2}. In summary, $\hat \Sigma_0$ is a compact manifold with boundaries, corners and
possibly conical singularities in the interior.
The quotient $\hat \H = \H/\K \subset \partial \hat \Sigma_0$ is represented by a boundary segment in the first case
described in thm.~\ref{thm2}, i.e. when the horizon topologies are as in eq.~\eqref{htopology}
with $s=D-1$. It is represented by the boundary of a removed disk from $\hat \Sigma_0$ in the second case described in the thm.~\ref{thm2}, i.e. when the horizon topology is $\H \cong \T^{D-1}$.
The other boundary components of $\hat \Sigma_0$ are line segments corresponding to axis. The quotients $\hat \Sigma_0$
and $\hat \Sigma_\infty$ are glued together along the joint boundary
$\{(R_1, R_2) \in \mr^2 \mid R_1, R_2 > 0, \,\, R_1^2 + R^2_2 =r^2\}$.
%, see figure~\ref{fig1}.
It is clear that $\Sigma_0$ is oriented and connected. Therefore, it must be a handle body with
possibly different boundary components, each homeomorphic to circles, and with conical singularities in the interior.
 Gluing $\hat \Sigma_\infty$
onto $\hat \Sigma_0$, we thus see that $\hat \Sigma$ is homeomorphic to the connected sum of a
half-plane and a handle body $\hat \Sigma_g$, with a number of disks removed and with orbifold points.
Therefore, topologically
\ben\label{hatsigmadecomp}
\hat \Sigma \cong (\mr \times \mr_{>0}) \,\, \# \,\, \hat \Sigma_g \, \setminus \bigcup_{j=1}^d D^2_j \, .
\een

To rule out the presence of handles, removed disks, and points with conical singularities,
we now use the topological censorship theorem for asymptotically Kaluza-Klein spaces~\cite{Chrusciel:2008uz},
see also~\cite{Woolgar,Woolgar1}.
This theorem states that any curve $\gamma$ with endpoints in $\Sigma_\infty$ can be continuously deformed to
a curve entirely within $\Sigma_\infty$. Furthermore, any closed loop in $\Sigma_\infty$ is
homotopic to a closed loop of in a neighborhood of $\Sigma_\infty$ of the form
$D^2 \times \K$, where $D^2$ is a two-dimensional disk that can be identified with a corresponding
disk in $\hat \Sigma_\infty$.
These facts together imply that if $f: D^2 \times \K  \to \Sigma$ is the
embedding map, then $f_*: \pi_1(D^2 \times \K) \to \pi_1(\Sigma)$ is surjective.
If $\Sigma$ were a compact manifold (without boundary), we could now directly apply lemma~\ref{lemmapione},
and thereby conclude can be no handles, removed disks, nor conical singularities,
and that $\hat \Sigma$ would consequently be a disk. In the case at hand, $\Sigma$ is a
manifold with boundary components consisting of $\H$ and ${\mathbb S}^3 \times \T^{D-5}$.
Nevertheless, using eq.~\eqref{hatsigmadecomp}, the arguments leading to lemma~\ref{lemmapione}
and~\ref{fundathm} can be very easily adapted to this case, or one may alternatively compactify
$\Sigma$ by gluing in appropriate manifolds with boundary $\H$ and ${\mathbb S}^3 \times \T^{D-5}$.
In either case, we conclude that $\hat \Sigma$ is a homeomorphic to a half-plane, and that there
are no conical singularities.

Since $\hat M = \langle \! \langle M \rangle \! \rangle/\G \cong \hat \Sigma$, this proves the theorem.
\qed

\subsection{Model spaces, examples}\label{subsec34}

We finally discuss to what extent the structure of the space $\langle\!\langle M \rangle\!\rangle$
as a manifold with $\G$-action is determined by the associated data described in
thm.~\ref{thm3}. As we have seen in the proof
of this theorem, the study of $\langle\!\langle M \rangle\!\rangle$
as a manifold with $\G$-action essentially boils down to the study of a $(D-1)$-dimensional spatial
slice $\Sigma$ with corresponding action of $\K$, and it is hence sufficient from a topological viewpoint
to study this situation.

Thus, let us assume that we are given an oriented $s$-dimensional
manifold $\Sigma$ with $\K$-action, with corresponding
orbit space $\hat \Sigma$ and decoration data, as described in prop.~\ref{proposition1}, and thm.~\ref{fundathm}.
For simplicity, let us consider the case that $\Sigma$ has no boundaries. The general case can be treated
quite similarly. Then we can ask whether $\Sigma$ as a manifold with $\K$-action is uniquely determined by the orbit space
and decoration data. In other words, given another such manifold $\Sigma'$, does there exist a diffeomorphism
$h: \Sigma \to \Sigma'$, and an automorphism $\alpha_A: \K \to \K$ such that $h(k \cdot x) = \alpha_A(k) \cdot h(x)$
for all $x \in \Sigma, k \in \K$? As shown in the case $s=4$ in~\cite[Para. I]{orlik2}, the answer
to this question is in the affirmative. (In the case that $\partial \hat \Sigma \neq 0$, the decoration
data must include also the invariant mentioned in remark~(2) after prop.~\ref{proposition1}.) The proof of this
theorem really extends straightforwardly to the case of $\Sigma$ with arbitrary dimension, so we will not describe it
here in detail.

A related question is whether for a given $\hat \Sigma$ and given decoration data as described in
prop.~\ref{proposition1}, we can find a corresponding manifold $\Sigma$ with $\K$-action described by
these data. The question is again in the affirmative, and we now outline how one can construct such
a manifold. Thus, let us assume that we are given (i) an orbit space $\hat \Sigma$ which is
an oriented two-dimensional manifold with boundaries, corners, and conical singularities,
(ii) vectors $\{ \underline a(I_j) \}$, one for each component $I_j \subset \partial \hat \Sigma$,
satisfying the constraints~\eqref{consistency} (iii) a collection $\{q_i, \underline p_i\}$,
one for each conical singularity $\hat x_i \in \hat \Sigma$, as described in around~\eqref{homoj}. We want to construct
a corresponding manifold $\Sigma$ with $\K$-action.

For simplicity, let us assume that $\hat \Sigma$ is a half-plane $\mr_{>0} \times \mr$, with
finitely many conical singularities in the interior, and with boundary divided into the
segments $I_1, \dots, I_b$. We first consider the conical singularities in the interior. We
may assume that they are all in a disk $D^2 \subset \mr_{>0} \times \mr$. We cut out this disk,
and we consider $D^2 \times \K$ with standard action of $\K$ on the second factor. We cut out
from this region $c$ tubes of the form $D_i^2 \times \K$, with each $D_i^2$ a small disk
containing the $i$-th of the $c$ conical singularities. Near the conical singularities, we
would like the $\K$-action to be described by the homomorphisms $\vartheta_i^{-1}: \Z_{q_i} \to \K$ given in
eq.~\eqref{homoj}. A model space for this action is
\ben
D_i^2 \times_{\vartheta^{-1}_i} \K \,\,\, , \quad D^2_i = \{ z \in \mc \mid  \,\, |z-z_i| \le 1 \} \, ,
\een
where $g \in \Z_{q_i} \subset {\mathbb S}^1$ acts on the disk by multiplication with the complex phase.
We glue in these model spaces along the boundaries where we cut out the $c$ tubes $D_i^2 \times \K$
with diffeomorphisms $h_i : \partial(D_i^2 \times_{\vartheta^{-1}_i} \K) \to \partial(D_i^2 \times \K)$
in such a way that the $\K$-actions match up. We call the manifold with boundary obtained from
$D^2 \times \K$ in this way $\Sigma_0$.

We now construct a second $\K$-space $\Sigma_1$ that incorporates the data $\{\underline a(I_j)\}$.
These data were constructed above
by giving, for each orbit, a neighborhood together with a set of coordinates in which the
action of $\K$ was explicitly given. It is intuitively clear that we can turn this around
and define $\Sigma_1$ to be the collection of these coordinate charts with corresponding $\K$-action,
and we now briefly explain how this can be done. For simplicity and concreteness,
we consider explicitly the case when $s=\dim \Sigma=4$. The construction is well-known in topology
and is sometimes called ``linear plumbing'', see~\cite{Hirzebruch}. We present the construction in
such a way that the generalization to general $s$ should be fairly obvious, details will be given
in~\cite{chrnew}.

The construction of $\Sigma_1$ is as follows. Let $b \ge 2$ be the number
of boundary segments $\{I_j\}$. On the boundary
$S^3$ of the four-dimensional solid ball $B^4 = \{
y_1^2+y_2^2+y_3^2+y_4^2 < 1\}$, we consider the disjoint subsets
\bena
S_+ &:=& \{ (y_1, y_2, y_3, y_4) \in {\mathbb S}^3 \mid \sqrt{y_3^2+y_4^2} < 1/4 \} \, ,\nonumber\\
\quad
S_- &:=& \{ (y_1, y_2, y_3, y_4) \in {\mathbb S}^3 \mid \sqrt{y_1^2+y_2^2} < 1/4 \} \, .
\eena
Both of these subsets are topologically solid tori. We consider the disjoint
union of $b-1$ copies of the solid ball $B^4$, and on the $i$-th copy we define
an action of $\K = \T^2$ generated by the two $2\pi$-periodic vector
fields $\psi_1, \psi_2$ given by
\ben\label{psiy}
\left(
\begin{matrix}
y_1 \partial_{y_2} - y_2 \partial_{y_1}\\
y_3 \partial_{y_4} - y_4 \partial_{y_3}
\end{matrix}
\right)
=
\left(
\begin{matrix}
a_1 (I_i) & a_2(I_i)\\
a_1 (I_{i+1}) & a_2(I_{i+1})
\end{matrix}
\right)
\left(
\begin{matrix}
\psi_1\\
\psi_2
\end{matrix}
\right)
%= A_i \left(
%\begin{matrix}
%\psi_1\\
%\psi_2
%\end{matrix}
%\right)
\, .
\een
The consistency condition on the $i$-th corner~\eqref{consistency1},~\eqref{consistency} guarantees that
the determinant of the above matrix is $\pm 1$.
We wish to glue the $S_+$-part of the boundary of the $i$-th copy of the
ball $B^4$ to the $S_-$-part of the boundary of the $(i+1)$-th
copy in such a way that the actions of $\K$ on
these copies are compatible. It is not difficult to
see that this is achieved if we identify these parts
by the maps $f_i: S_- \to S_+$
defined by
\ben
f_i(y_1, y_2, y_3, y_4) = \Big(
y_3, y_4, y_1 \sin (n_i\varphi) + y_2 \cos (n_i\varphi),
y_1 \cos (n_i\varphi) - y_2 \sin (n_i\varphi)
\Big) \, ,
\een
where $\varphi = \arctan \frac{y_3}{y_4}$ and $n_i = a_1(I_i) a_2(I_{i+2})-a_2(I_i)a_1(I_{i+2})$,
i.e. we have $f_{i \, *} \psi_1 = \psi_1$ and $f_{i \, *} \psi_2 = \psi_2$. Thus, for $b>2$ we define\footnote{
If $X,Y$ are sets and $f$ is a map $f: A \subset X \to Y$, then $X \cup_f Y$ is the set defined
as the quotient of the disjoint union $X \cup Y$ by the equivalence relation
$x \sim y : \Leftrightarrow (x,y) \in {\rm graph}\, f$.
}
\ben\label{Xdef}
\Sigma_1 = (\dots ((B^4 \cup_{f_1} B^4) \cup_{f_2} B^4) \dots \cup_{f_{b-3}} B^4) \cup_{f_{b-2}} B^4 \, .
\een
For $b=2$ be define $\Sigma_1 = B^4$. The space $\Sigma_1$ has a 3-dimensional boundary whose structure is
determined by the first and last vector $\underline a(I_1)$, and $\underline a(I_b)$. It is either
$\T^1 \times {\mathbb S}^2, {\mathbb S}^3$, or a lens space $L(p,q)$, see thm.~\ref{thm2}.

We may cut out from $\Sigma_1$ a tube $D^2 \times \K$, and glue the boundary obtained in
this way onto $\partial \Sigma_0$. The manifold $\Sigma$ obtained in this way is
the desired $\K$-space $\Sigma$ in the special case considered. The general case
may be treated in a similar way, as we will discuss in a future paper~\cite{chrnew}.
We may call the manifold $\Sigma$ constructed from the decoration data of the orbit space
$X[\epsilon, \hat \Sigma, \{\underline a(I_i)\}, \{q_i, \underline p_i\}]$, where
$\epsilon$ is an orientation, and $\hat \Sigma$ an oriented two-dimensional manifold with
boundaries and corners. We give some examples (without conical singularities):

\paragraph{Example 1:} (From~\cite{orlik1}) Let $s=4$, $\hat \Sigma = D^2, \partial D^2 = I_1 \cup I_2 \cup I_3$,
and consider the data $\{ (1,0), (0,1), (1,1) \}$. Then
the space $X[D^2, \{(1,0),(0,1),(1,1)\}]$ is the complex projective space $\mc P^2 = \mc^3/\sim$, where the equivalence
relation is $(z_1, z_2, z_3) \sim (\lambda z_1, \lambda z_2, \lambda z_3)$ and the action of $\K = \T^2$ is $
[\tau_1, \tau_2] \cdot (z_1, z_2, z_3)_\sim = (\e^{i\tau_1} z_1, \e^{i\tau_2} z_2, z_3)_\sim$. The
equivalence $X[D^2, \{(1,0),(0,1),(1,1)\}] \cong \mc P^2$ can be seen e.g.
by noting that the axis in $\mc P^2$ corresponding to the vectors $(1,0), (0,1), (1,1)$ are given by
the set of points $(z_1, z_2, z_3)_\sim \in \mc P^2$ such that, respectively, $z_1 = 0$, $z_2 = 0$, $z_3 = 0$.

\paragraph{Example 2:} Let $s=4$, $\hat \Sigma = D^2$ and consider the data $\{ (1,0), (0,1), (1,0), (0,1) \}$ (four intervals). Then
the space $X[D^2, \{(1,0),(0,1),(1,0),(0,1)\}]$ is ${\mathbb S}^2 \times {\mathbb S}^2$, with the standard action of $\K$.
This is easily seen by considering the isotropy groups of the action. In fact, examples~1 and~2 constitute in some sense the
most general case in $s=4$ because one can show that~\cite{orlik1,orlik2},
topologically, $\Sigma$ is a connected sum of projective spaces an
${\mathbb S}^2 \times {\mathbb S}^2$'s in the situation under consideration.

\paragraph{Example 3:} Let $s=5$, $\hat \Sigma = D^2$ and consider the data $\{(1,0,0), (q_1, q_2, p), (0,1,0)\}$.
The constraints on the corners are fulfilled if we have ${\rm g.c.d.}(p,q_1) = 1 = {\rm g.c.d.}(p,q_2)$.
The corresponding space $X[D^2, \{(1,0,0), (q_1, q_2, p), (0,1,0)\}]$ is a {\em generalized lens space}
$L(p; q_1, q_2)$. The generalized lens space is defined as the quotient of ${\mathbb S}^5$
(realized as the unit sphere in $\mc^3$) by the discrete subgroup of isometries of order $p$ generated by an element
$\lambda$ acting as $\lambda \cdot (z_1, z_2, z_3) = (\e^{2\pi i/p} z_1, \e^{2 \pi i q_1/p} z_2, \e^{2 \pi i q_2/p} z_3)$.
The action of $\K = \T^3$ on an equivalence class $(z_1, z_2, z_3)_\sim \in L(p; q_1, q_2)$ under this action is
\ben
[\tau_1, \tau_2, \tau_3] \cdot (z_1, z_2, z_3)_\sim = (\e^{i \tau_3/p} z_1, \e^{i(\tau_1 + q_1 \tau_3/p)} z_2,
\e^{i(\tau_2 + q_2 \tau_3/p)} z_3)_\sim \, .
\een
The axis corresponding to the vectors $(1,0,0), (q_1, q_2, p), (0,1,0)$ are, respectively,
$z_2 = 0, z_2 = z_3 = 0 , z_3= 0$. Note that $\pi_1 (L(p; q_1, q_2)) \cong \Z_p$, so for $p \neq 1$  this
space is not simply connected.

\paragraph{Example 4:} Let $s,\hat \Sigma$ be as in the previous example, but let the data now be
$\{(1,0,0), (q_1, q_2, p), (0,1,0), (1, 1, 0)\}$.
The constraints on the corners are fulfilled if we have ${\rm g.c.d.}(p,q_1) = 1 = {\rm g.c.d.}(p,q_2)$. The manifold in
question is now topologically (combining the examples~1 and~3)
\ben
X[D^2, \{(1,0,0), (q_1, q_2, p), (0,1,0), (1, 1, 0)\}] \cong L(p; q_1, q_2) \# (\mc P^2 \times {\mathbb S}^1) \, .
\een

\section{Stationary vacuum black holes in $D$ dimensions}
\label{sec3}

In the previous section, we looked at the topology of the domain of outer communication $\langle\!\langle M \rangle\!\rangle$ and the
structure of the orbits of the symmetries. In this section, we investigate the spacetime metric, i.e. the
implications of the Einstein equations $R_{ab}=0$.

These equations imply a set of coupled
differential equations for the metric on the two-dimensional factor space $\hat M$,
described above in thm.~\ref{thm3}.
To understand these equations in a geometrical way, we note that
the projection $\pi: \langle\!\langle M \rangle\!\rangle \to
\langle\!\langle M \rangle\!\rangle/\G=\hat M$
(with $\G = \T^{D-3} \times \mr$ the isometry group) defines a $\G$-principal fibre
bundle over the interior of $\hat M$, because we argued in the previous section
that such points correspond to points in the domain of outer communication
with trivial isotropy group. At each point $x \in \langle\!\langle M \rangle\!\rangle$ in a fibre over $\pi(x)$ in the interior of
$\hat M$, we may uniquely decompose the tangent space at $x$ into
a subspace of vectors tangent to the fibres, and a space $W_x$ of vectors
orthogonal to the fibres. Evidently, the distribution of vector spaces
$W_x$ is invariant under the group $\G$ of symmetries, and hence forms
a ``horizontal bundle'' in the terminology of principal fibre
bundles~\cite{Kobayashi}. According to standard results in the theory of principal
fibre bundles~\cite{Kobayashi}, a horizontal bundle
is equivalent to the
specification of a $\G$-gauge connection $\hat D$ on the factor
space, whose curvature we denote by $\hat F = T_I \hat F^I_{\alpha\beta} dx^\alpha \wedge dx^\beta$,
with $T_I, I=0, \dots, D-3$ the generators of the abelian group $\G$. Roman indices $\alpha,\beta, \dots$ take the values
$1,2$. The horizontal
bundle gives an
isomorphism $W_x \to T_{\pi(x)} \hat M$ for any $x$, and
this isomorphism may be used to uniquely construct a smooth
covariant tensor field
$\hat t_{\alpha\beta \dots \gamma}$ on the interior of $\hat M$
from any smooth $\G$-invariant
covariant tensor field $t_{ab \dots c}$
on $M$.

For example, the metric $g_{ab}$ on $M$ thereby gives rise to
a symmetric tensor $\hat g_{\alpha\beta}$ on $\hat M$. One can show with a
significant amount of labor~\cite{chrnew} (see also \cite{chrcos}) that the $D-2$ dimensional subspaces
spanned by the Killing fields at points of $\langle\!\langle M \rangle\!\rangle$
corresponding to interior points of $\hat M$ always contain a timelike vector.
Hence the bilinear form induced from $g_{ab}$ on $W_x$
has signature $(++)$, so $\hat g_{\alpha\beta}$ is in fact a Riemannian metric. We let $\hat D$
act on ordinary tensors $\hat t_{\alpha\beta \dots \gamma}$ as the connection of
$\hat g_{\alpha\beta}$, with Ricci tensor denoted $\hat R_{\alpha\beta}$.

By performing the well-known Kaluza-Klein reduction of the metric
$g_{ab}$ along the orbits of $\G$, we can locally write the Einstein equations as a system of
equations on the interior of the factor space
$\hat M$ in terms of metric $\hat g_{\alpha\beta}$, the components
$\hat F_{\alpha\beta}^I$
of the curvature, and the  Gram matrix
field $G_{IJ}$
\begin{equation}\label{gramdef}
G_{IJ} = g(X_I, X_J) \, , \quad
X_I =
\begin{cases}
t & \text{if $I=0$}, \\
\psi_{i} & \text{if $I=i=1, \dots, D-3$.}
\end{cases}
\end{equation}
The resulting equations are similar in nature to the ``Einstein-equations''
on $\hat M$ for $\hat g_{\alpha\beta}$, coupled to the ``Maxwell fields''
$\hat F_{\alpha\beta}^I$ and
the ``scalar fields'' $G_{IJ}$, see~\cite{KK1,KK2}.
 We will not write these equations down
here, as we will not need them in this most general form.

In our case, the equations simplify considerably because one can show (see e.g.~\cite{chrcos}) that
the distribution of horizontal subspaces $W_x$ is locally
integrable, i.e., locally tangent to a family of
two-dimensional submanifolds.
In that case, the connection is flat,
$\hat F_{\alpha\beta}^I = 0$, and the dimensionally
reduced equations may be written as
\begin{equation}\label{reduced2}
\hat D^\alpha ( r G^{-1} \hat D_\alpha G)= 0 \, ,
\end{equation}
together with
\begin{equation}\label{reduced3}
\hat R_{\alpha\beta} = \hat D_\alpha \hat D_\beta \log r
-\frac{1}{4} {\rm Tr}\Big( \hat D_\alpha G^{-1} \hat D_\beta G
\Big) \, .
\end{equation}
Greek indices have been raised with $\hat g^{\alpha\beta}$.
The equations are well-defined a priori only at
points in the interior of $\hat M$ where the Gram determinant
\begin{equation}
r^2 = -{\rm det} \, G
\end{equation}
does not vanish.
Chrusciel has shown~\cite{chrnew}
(based on previous work of Carter~\cite{Carter} and also of~\cite{chrcos}) that
$r^2>0$ away from the boundary of $\hat M$.
%Their argument
%relies at one step on the fact that $\langle \! \langle M \rangle \! \rangle$ is
%simply connected when $D=4$, which does not hold for the asymptotically
%Kaluza-Klein spacetimes under consideration here. However, in our situation it is still
%the case~\cite{Chrusciel:2008uz} that any curve with endpoints in the asymptotic region
%can be deformed to a curve that is entirely within the asymptotic region. This appears to be
%enough for the argument of~\cite{chrcos} to work also in our case, showing
%that $r$ is a smooth and well defined positive function on the interior of $\hat M$, and
The reduced Einstein equations are hence well-defined there.
On the other hand, $r$ vanishes on any boundary component $I_j$ of $\hat M$ corresponding to an axis, i.e.
where a linear combination $\sum a_i(I_j) \psi_i = 0$ vanishes, because the Gram matrix then
has a non-trivial kernel. It also vanishes on the segment of $\partial \hat M$ corresponding
to the horizon $H$, because the span of $X_I, I=0, \dots, D-3$ is tangent to $H$
and hence a null space, with the signature of $G$ consequently being $(0++\dots+)$ there.

Taking the trace of the first reduced Einstein equation~\eqref{reduced2},
one finds that $r$ is a harmonic function on the interior of
$\hat M$,
\begin{equation}
\hat D^\alpha \hat D_\alpha r = 0 \, .
\end{equation}
Since $\hat M$ is an (orientable) simply connected 2-dimensional
analytic manifold with connected boundary and corners by thm.~\ref{thm1},
we may map it analytically to the upper complex
half plane $\{\zeta \in \mc \mid \,\, {\rm Im} \, \zeta > 0\}$ by the
Riemann mapping theorem. Furthermore, since $r$ is harmonic, we can
introduce a harmonic scalar field $z$ conjugate to
$r$
\ben\label{zdef}
\hat D^\alpha z = \hat \epsilon^{\alpha\beta} \hat D_\beta r \, ,
\een
where $\hat \epsilon_{\alpha\beta}$ is the  anti-symmetric tensor on $\hat M$
satisfying $\hat \epsilon^{\alpha\beta} \hat \epsilon_{\alpha\beta} = 2$.
Thus both $r,z$ are harmonic functions on $(\hat M, \hat g)$, and
$r = 0$ on $\partial \hat M$. Combining this with the fact $\hat M$ is homeomorphic to
a half-plane, one can argue (see e.g.~\cite[6.3]{chrcos} or~\cite{Weinstein}) that
%Since an analytic mapping is conformal we also have
%$\partial_\zeta \partial_{\bar \zeta} r = 0 =
%\partial_\zeta \partial_{\bar \zeta} z$, and from this, together with
%the boundary condition $r=0$ for ${\rm Im} \, \zeta = 0$, one can
%argue that $\zeta = z + ir$ by a simple argument involving the maximum
%principle~\cite{Weinstein}. In particular,
$r$ and $z$ are globally defined coordinates, and identify $\hat M$ with
$\{ z + ir \in \mc \mid
r> 0\}$. In these coordinates, the metric $\hat g$ globally takes the form
\begin{equation}\label{hatg}
\hat g = \e^{2\nu(r,z)} (\d r^2 + \d z^2) \, .
\end{equation}
Since~eq.\eqref{reduced2} is invariant under conformal rescalings
of $\hat g_{\alpha\beta}$, and since a 2-dimensional metric is conformally
flat, it decouples from eq.~\eqref{reduced3}. In fact, writing the
Ricci tensor $\hat R_{\alpha\beta}$ of~\eqref{hatg}
in terms of $\nu$, one sees that eq.~\eqref{reduced3}
equation may be used to determine $\nu$ by a simple integration, see
e.g.~\cite{Harmark} for details.

The boundary $r=0$ of $\hat M$  consists of several
segments according to our classification theorem~\ref{thm3}. In the description of
$\hat M$ as the upper complex half plane $\hat M = \{z+ir \in \mc \mid r>0\}$, these
are represented by a collection of intervals $\{I_j\}$ of the $z$-axis. The length of
the $j$-th interval as measured by the coordinate $z$ is called $l(I_j)$. Because the coordinates $(r,z)$ were canonically
defined, the numbers $l(I_j) \ge 0$ are invariantly defined, i.e. are the same
for isometric spacetimes. Each segment is either an
axis for which there is a vector $\underline a(I_j) \in \Z^{D-3}$
such that $\sum_i a_i(I_j) \psi_i=0$, or it corresponds to the horizon. In that case,
we put the corresponding vector to zero, $\underline a_H = 0$, because no non-trivial
linear combination of the axial Killing fields vanishes in the interior of the
corresponding interval $I_H$, see thm.~\ref{thm3}. Concerning the length $l_H$ of the
horizon segment, we have the following lemma.
\begin{lemma}\label{lemma6}
The length of the horizon interval satisfies
\ben
(2\pi)^{D-3} l_H = \kappa A_H \,\,\,  ,
\een
where $A_H$ is the area of the horizon cross section $\H$, and where
$\kappa>0$ is the surface gravity.
\end{lemma}
The proof of lemma~\ref{lemma6} is given in appendix~\ref{appendixa}.

\medskip
\noindent
We call the collection of real positive numbers $\{l(I_j)\}$ and integer vectors $\{\underline a(I_j)\}$
associated with the intervals the ``interval structure'' of the spacetime. As we explained in the
previous section, the collection $\{\underline a(I_j)\}$ determines the
manifold structure of $\langle\!\langle M \rangle \! \rangle$ and the action of $\G$ on this space
up to diffeomorphism. In particular, the vector fields $X_I$ are determined
up to diffeomorphism. Furthermore, if we are given
$G_{IJ}$ and $\hat g$ (i.e., $\nu$) as functions of $r, z$, then we can reconstruct the metric $g$
of the spacetime in the domain of outer communication.
In a local coordinate system consisting of $r,z$ and
$\xi^I, I=0, \dots, D-3$, such that the
Killing fields are given by $X_I = \partial/\partial \xi^I$, the metric locally
takes the form
\ben\label{metric}
g = \e^{2\nu(r,z)}(\d r^2 + \d z^2) + G_{IJ}(r,z) \, \d\xi^I \d\xi^J \, .
\een
For $M=\mr^{4,1} \times \T^{D-5}$, the axial symmetries
are the rotations in the $12$-plane of $\mr^{4,1}$ generated by the Killing field
$\psi_1$, the rotations in the
the $34$-plane of $\mr^{4,1}$ generated by the Killing field $\psi_2$
and the rotations of the $D-5$ compact extra dimensions generated by Killing fields $\psi_3, \dots, \psi_{D-3}$.
The coordinates $r,z$ as constructed above
are given by
$r=R_1 R_2$ and $z = \frac{1}{2}(R_1^2-R_2^2)$, with $R_1 =
\sqrt{x_1^2+x^2_2}$ and $R_2 = \sqrt{x_3^2+x^2_4}$, and with
$x_i$ the standard spatial Cartesian coordinates of
$\mr^{4,1}$. The conformal
factor is given by $\e^{2\nu} = 1/2\sqrt{r^2+z^2}$, and the Gram matrix is
given by
\ben
G =
\left(
\begin{matrix}
-1 & 0 & 0 & 0 & \dots & 0\\
0 & \rho(1-\cos \theta) & 0 & 0 & \dots & 0\\
0 & 0 & \rho(1+\cos \theta) & 0 & \dots & 0\\
0 & 0 & 0                   & 1 & \dots & 0\\
\vdots & \vdots & \vdots & \vdots & & \vdots \\
0 & 0 & 0 & 0 & \dots & 1
\end{matrix}
\right) \, .
\een
Here, we have introduced the coordinates $\rho, \theta$ which are related to $r,z$ by
\ben
r=\rho \sin \theta \, , \quad z = \rho \cos \theta  \, ,
\een
or
\ben
\rho = \frac{1}{2}(x_1^2+x_2^2+x_3^2+x_4^2) \, , \quad
\theta = \arctan \frac{2\sqrt{(x_1^2+x_2^2)(x_3^2+x_4^2)}}{x_1^2-x_3^2+x_2^2-x_4^2} \,
\een
in terms of the spatial cartesian coordinates $x_i$ of $\mr^{4,1}$. For a general
$D$-dimensional asymptotically Kaluza-Klein spacetime
with asymptotically flat 5-dimensional part, we can determine the asymptotic form of the metric as
follows: First, one establishes, using standard results on elliptic equations,
that $G$ has a poly-homogeneous asymptotic expansion in powers of
$1/\rho$ for large $\rho$ of the form
\ben\label{gramasympt}
G(\rho,\theta) \sim \sum_{n \ge 0} \rho^{-n} G_{n} (\rho,\theta) \, ,
\een
where $G_0$ is the diagonal Gram matrix for $\mr^{4,1} \times \T^{D-5}$ given
above, and where the other terms $G_n$ represent corrections. The entries of the correction matrices
are of the same order in $\rho$ as those of $G_0$, up to possibly additional powers of $\log \rho$.
We then insert this ansatz into the first reduced Einstein equation eq.~\eqref{reduced2}.
Since the leading part $G_0$ is a solution
to the equation, we get from this an equation for the correction matrix elements.
At the lowest non-trivial order in $1/\rho$ this equation delivers a decoupled system of second order
ordinary differential equations in $\theta$ for the entries of the correction matrix $G_1$.
These equations have a unique solution satisfying the boundary conditions arising from
the fact that the $11,12,13,\dots$-components of $G$ must vanish for $\theta=0$ and
sufficiently large $\rho$ (as this represents an axis for $\psi_1$), while the $21,22,23,\dots$-components must vanish for $\theta=\pi$
and sufficiently large $\rho$ (as this represents an axis for $\psi_2$). Hence, these components must similarly vanish also for $G_1$.
We do not give the details of the straightforward but
somewhat lengthy calculation but only quote the solution, written in
a block-matrix form:
\ben
G_1 =
\left(
\begin{matrix}
2M & b_1 (1-\cos \theta) & b_2 (1+ \cos \theta) & (b_i) \\
b_1(1-\cos \theta) & (M-A+\eta)(1-\cos \theta) \rho   & \zeta \sin^2 \theta & (c_i)(1-\cos \theta) \\
b_2(1+\cos \theta) &  \zeta  \sin^2 \theta & (M-A-\eta)(1+\cos \theta) \rho & (d_i)(1+\cos \theta) \\
(b_i) & (c_i)(1-\cos \theta) & (d_i)(1+\cos \theta) & (h_{ij})
\end{matrix}
\right) \, .\nonumber
\een
Here, the quantities
$M, A, \zeta, \eta, h_{ij}, b_i, c_i, d_i$ are undetermined real constants and $i,j$ range through $3, \dots, D-3$ in this
block-matrix. Because we must have
$-{\rm det} \, G = r^2$, they are subject to the constraint
$$2A=\sum_{i=3}^{D-3} h_{ii} \, .$$
According to eq.~\eqref{zdef}, we are still free to change the coordinate $z$ by
adding a constant. This will result in adding a constant to $\eta$, and we may thus
fix the remaining ambiguity in $z$ by setting $\eta = 0$. We will do this in the following.
The asymptotic form of the conformal factor $\e^{2\nu}$ can similarly be determined by
the second reduced Einstein equation, eq.~\eqref{reduced3}, together with
the asymptotic form of the Gram matrix eq.~\eqref{gramasympt}. Again, we omit the
straightforward but somewhat lengthy calculation and give only the result, which is
\ben
\e^{2\nu} = \frac{1}{2\rho} + \frac{M-A}{4\rho^2} + \dots\, ,
\een
where the dots represent terms that go to zero faster as $\rho \to \infty$.
Thus, in a coordinate system $(\tau, \rho, \theta,
\varphi_1, \dots, \varphi_{D-3})$ such that
\ben
t = \partial/\partial \tau \, , \quad \psi_i = \partial/\partial \varphi_i \, , \quad i=1, \dots, D-3 \, ,
\een
we obtain the following asymptotic form of the metric eq.~\eqref{metric} for large $\rho$:

\medskip
\paragraph{Asymptotic form of the metric} for stationary black hole spacetime with
$D-3$ axial Killing fields, behaving as
$\mr^{4,1} \times \T^{D-5}$ near infinity:
\bena\label{metricinfty}
g &=& -\left(1 - \frac{2M}{\rho} \right) \d\tau^2 + \frac{1}{2\rho}\left(
1+\frac{M-A}{2\rho}
\right) (\d\rho^2 + \rho^2 \d\theta^2) \nonumber\\
&&+\rho(1-\cos \theta)\left(1+ \frac{M-A}{\rho} \right)
\d\varphi_1^2
+ \rho(1+\cos \theta)\left(1+ \frac{M-A}{\rho} \right)\d\varphi_2^2 \nonumber\\
&&+ \sum_{i,j=3}^{D-3} \left(\delta_{ij} + \frac{h_{ij}}{\rho} \right) \d\varphi_i \d\varphi_j
+ \frac{2\zeta \sin^2 \theta}{\rho} \, \d\varphi_1 \d\varphi_2 \nonumber\\
&&+ \frac{2b_1(1-\cos \theta)}{\rho} \, \d\varphi_1 \d\tau
+ \frac{2b_2(1+\cos \theta)}{\rho} \, \d\varphi_2 \d\tau + \frac{2}{\rho} \sum_{i=3}^{D-3} b_i \, \d\varphi_i \d\tau \nonumber\\
&&+ \frac{2(1 + \cos \theta)}{\rho} \sum_{i=3}^{D-3} d_{i} \, \d\varphi_2 \d\varphi_i +
\frac{2(1 - \cos \theta)}{\rho} \sum_{i=3}^{D-3} c_{i} \, \d\varphi_1 \d\varphi_i + \dots \, ,
\eena
where the dots represent terms that are
higher order in $1/\rho$.
The constants $b_i$ are proportional  to the angular momenta of the solutions, both in the asymptotically small
and large dimensions. These can be defined e.g. by the Komar expressions
\ben
%m = -\frac{3}{2} \int_{\S^3 \times \T^{D-5}} *\d t \, , \quad
J_i = \int_{\S^3 \times \T^{D-5}} *\d\psi_i  \, ,
\een
where the integration is over a surface at infinity, and where $\d\psi_i$ denote the 2-forms
obtained by taking the exterior differential of $\psi_i$ after lowering the index.
The constant $M$ is related to the ADM-mass of the solution, see e.g.~\cite[sec.~3]{traschen}.

\medskip
For $M = \mr^{3,1} \times \T^{D-4}$,
the axial symmetries may be taken as the rotations in the $12$-plane of $\mr^{3,1}$ and rotations of the
$D-4$ compact extra dimensions. The functions $r,z$ are then
given by $r=\sqrt{x_1^2+x_2^2}$ and $z = x_3$,
with $x_i$ the standard spatial Cartesian coordinates
on $\mr^{3,1}$. The conformal factor is just $\e^{2\nu} = 1$. For a general
$D$ dimensional asymptotically Kaluza-Klein spacetime
with asymptotically flat 4-dimensional part, we may again derive an expression for the asymptotic form of the metric as above. The same can also be done for $M=\mr^{2,1} \times \T^{D-3}$ and
$M= \mr^{1,1} \times \T^{D-2}$.
Since the analysis is quite similar, we do not give the results here.

\section{Uniqueness theorem for stationary black holes with $(D-3)$ axial symmetries}
\label{sec4}
In the previous two sections, we have analyzed stationary black hole spacetimes
that are asymptotically $\mr^{s,1} \times \T^{D-s-1}$ where $s=1,2,3$ or $4$, and
which have an isometry group $\G = \mr \times \T^{D-3}$ (with no points in the domain of
outer communication whose isotropy group is discrete). We have derived a number of
``invariants'' associated with such solutions:
\begin{itemize}
\item We showed that the orbit space of the domain of outer communication by $\G$ is
a half plane $\hat M = \{z + ir \mid r > 0\}$. The boundary of the half-plane is
divided into a finite collection of intervals $\{I_j\}$. With each interval, there
is associated its length\footnote{For a half infinite interval, this would be $\infty$.}
$l(I_j) \in \mr_{>0}$, and a vector $\underline a(I_j) \in \Z^{D-3}$ subject
to the normalization~\eqref{normalization}.
One of the intervals corresponds to the orbit space $\hat \H$ of the horizon and is associated with the
zero vector, while the others
correspond to an ``axis'' in spacetime, i.e. points
where the linear combination $\sum_i a_i(I_j) \psi_i = 0$ vanishes.
For adjacent intervals $I_j$ and $I_{j+1}$ (not including the horizon), there is a compatibility condition stating that
the collection of minors $Q_{kl} \in \Z, \,\, 1 \le k < l \le D-3$ given by
\ben
%\label{consistency1}
Q_{kl} = |\det \left(
\begin{matrix}
a_k(I_{j+1}) & a_k(I_j)\\
a_l(I_{j+1}) & a_l(I_j)
\end{matrix}
\right)| \,
\een
have greatest common divisor ${\rm g.c.d.}\{ Q_{kl} \}=1$, see the discussion around~\eqref{consistency}.
The data $\{l(I_j)\}$ together with $\{\underline a (I_j)\}$ were called
the ``interval structure''.
\item Because the spacetime is asymptotically Kaluza-Klein, we can define its mass, and
the angular momenta $\{J_i\}$ corresponding to the axial Killing fields, $i=1, \dots, D-3$.
%By convention, for spacetimes asymptotically to $\mr^{4,1} \times \T^{D-5}$, the first
%two $J_1, J_2$ correspond to the angular momentum in the large dimensions, while for
%spacetimes asymptotic to $\mr^{3,1} \times \T^{D-4}$, $J_1$ corresponds to angular
Some of the angular momenta correspond to the large, and some to the small
(extra) dimensions.
\item The asymptotic form of the metric~\eqref{metricinfty} contains additional real parameters
$\{h_{ij}\}, \{c_i\}, \{d_i\}, \zeta$ which are related to the asymptotic metric on the tori generated by the
axial Killing fields $\psi_i, i=1, \dots, D-3$ in the region of spacetime near infinity.
These numbers are invariantly defined.
%, up
%to $SL(\Z, D-3)$ transformations of the matrix elements $h_{ij} \to \sum A^k_i A^l_j h_{kl}$
%and a corresponding transformation of $c_i$ and $d_i$.
\item The collection of angular velocities $\{\Omega_i\}$, the surface gravity $\kappa$, and horizon area.
\end{itemize}
It is natural to ask the following questions: Is the spacetime $(M,g)$ under consideration uniquely determined
by the above data? To what extent can the data be specified independently?
The following theorem provides an answer to the first question and a partial answer to the second
question.
\begin{thm}\label{thm4}
There can be at most one stationary, asymptotically Kaluza-Klein spacetime $(M,g)$
with $D-3$ axial Killing fields, satisfying the technical assumptions stated in sec.~\ref{sec1},
for a given interval structure $\{ \underline a(I_j), l(I_j) \}$ and a given
set of angular momenta $\{J_i\}, i=1, \dots, D-3$.
\end{thm}
This uniqueness theorem is the main result of this paper. A consequence of the theorem is that
the interval structure and angular momenta uniquely determine the other invariants mentioned above,
such as e.g. the mass of the spacetime. In $D=4$ with no extra dimensions, the only non-trivial interval structure for a single black hole spacetime is given by the intervals $(-\infty, -z_0],
[-z_0, z_0], [z_0, \infty)$. The middle interval corresponds to the horizon, while the
half-infinite ones to the axis of the rotational Killing field. The interval vectors $\underline a(I_j)$
are 1-dimensional integer vectors in this case and hence trivial. For each $z_0>0$ and for each
angular momentum $J$, there exist a solution given by the appropriate member of the Kerr-family of
metrics.
Thus, the Kerr metrics exhaust all possible stationary, axially symmetric single black hole
spacetimes (satisfying the technical assumptions stated in sec.~\ref{sec1}). This is of course
just the classical uniqueness theorem for the Kerr-solution~\cite{Bunting,Carter,Mazur,Robinson,Hawking}, see~\cite{chrcos} for a rigorous account. The mass $m$ of the non-extremal Kerr solution characterized
by $z_0, J$ is related to these parameters by $z_0=\sqrt{m^2-J^2/m^2}>0$. Hence the uniqueness
theorem may be stated equivalently in terms of $m$ and $J$, which is more commonly done. Note that the
length of the horizon interval, $l_H = 2z_0$ tends to zero in the extremal limit, in accordance with
lemma~\ref{lemma6}.

In higher dimensions, one may similarly derive relations between the interval structure and angular
momenta on the one side, and the other invariants on the other side for any
given solution. Such formulae are provided for the Myers-Perry or black-ring solutions e.g. in~\cite{Harmark}. Of course, for most interval structures it is not known whether there actually
exists a solution, so in this sense much less is known in higher dimensions than in $D=4$.

\medskip
\noindent
{\em Proof of thm.~\ref{thm4}:} For definiteness, we give a proof here for spacetimes asymptotic to $\mr^{4,1} \times \T^{D-5}$,
the other cases are similar. We will show that the the domains of outer communication of any two spacetimes
as in the theorem must be isometric. It then follows from the argument given in~\cite{Friedrich} based on the
characteristic initial value formulation of the Einstein equations that the metrics of the interior of the two
black holes must also coincide. (The last step can be avoided if one assumes that the spacetime metric
is analytic.)

The key step is to define from the reduced Einstein
equations~\eqref{reduced2} a set of equations which
describe the difference between two solutions as described in the theorem. This formulation is due to~\cite{Mazur,Maison},
see also~\cite{Ida}, and it involves certain potentials which we define first.
We first consider the twist 1-forms
\ben
\omega_{i} = *(\psi_1 \wedge \dots \wedge \psi_{D-3} \wedge \d\psi_i) \quad i=1, \dots, D-3 \, ,
\een
where the Killing fields have been identified with 1-forms via the metric.
Using the vacuum field equations and standard identities for
Killing fields~\cite{Waldbook}, one shows that these 1-forms are closed, $\d\omega_i=0$.
Since the Killing fields commute, the twist forms are invariant under
$\G$, and so we may define corresponding 1-forms $\hat \omega_i$
on the interior of the factor space
$\hat M = \{z+ir \in \mc \mid r > 0\}$.
These 1-forms are again closed. Thus, the ``twist potentials''
\begin{equation}\label{lineint}
\chi_i = \int_0^{\hat x} \hat \omega_i
\end{equation}
are globally defined on $\hat M$ and independent of the path connecting
$0$ and the point $\hat x \in \hat M$, and $\d\chi_i = \hat \omega_i$.
The twist potentials and the
Gram matrix of the axial Killing fields $f_{ij} = g(\psi_i, \psi_j)$,
satisfy a system of coupled differential equations on $\hat M$ which
follow from the reduced Einstein equation~\eqref{reduced2}. They
are
\bena\label{ernst}
0&=&\hat D^\alpha \Big( r (\det f)^{-1} \chi^i \hat D_\alpha \chi_i + r \hat D_\alpha \log \det f\Big)\\
0&=&\hat D^\alpha \Big( r (\det f)^{-1} f^{ij} \hat D_\alpha \chi_j\Big)\\
0&=&\hat D^\alpha \Big( r f^{jk} \hat D_\alpha f_{ki} + r (\det f)^{-1} f^{jk} \chi_i \hat D_\alpha \chi_k\Big)\\
0&=&\hat D^\alpha \Big( -r\hat D_\alpha \chi_i + r \chi_i \hat D_\alpha \log \det f + r(f^{jk} \hat D_\alpha f_{ij}) \chi_k+\nonumber\\
&& \qquad r(\det f)^{-1} \chi^j (\hat D_\alpha \chi_j) \chi_i\Big) \, .
\eena
Here we are using the summation convention and $f^{ij}$ denotes the components of the inverse of
the matrix $f_{ij}$, which is used to raise indices on $\chi_i$. To verify these equations,
it is necessary to use the relations
\begin{equation}\label{Dalpha}
\hat D_\alpha \alpha^i = r(\det f)^{-1} \, \hat \epsilon_\alpha{}^\beta
\, f^{ij} \hat D_\beta \chi_j \, ,
\end{equation}
as well as
\begin{equation}\label{alphabeta}
\beta = f^{ij} \alpha_i \alpha_j - (\det f)^{-1} r^2
\end{equation}
for the scalar products $\alpha_i = g(t, \psi_i)$ and $\beta = g(t,t)$. Again,
$\alpha^i$ means $f^{ij} \alpha_j$. The above equations can be written in
a compact matrix form. For this, one introduces the $(D-2) \times (D-2)$
matrix field $\Phi$ which is written in an obvious block-matrix notation as
\begin{eqnarray}
\Phi=
\left(
  \begin{array}{cc}
    (\det f)^{-1} & -(\det f)^{-1} \underline \chi^T\\
    -(\det f)^{-1} \underline \chi & f + (\det f)^{-1} \underline \chi \otimes \underline \chi^T
  \end{array}
\right) \, ,
\end{eqnarray}
with $\underline \chi^T = (\chi_1, \dots, \chi_{D-3})$.
The matrix $\Phi$ satisfies $\Phi^T=\Phi$, $\det \, \Phi=1$, and is positive
semi-definite, being the sum of two positive semi-definite matrices. Hence
it may be written in the form
$\Phi= S^T S$ for some matrix $S$ of determinant 1. The
equations~\eqref{ernst} can be stated in terms of $\Phi$ as
\begin{eqnarray}\label{divergenceid}
\quad \quad
\hat D^\alpha (r \, \Phi^{-1} \hat D_\alpha \Phi) =0 \, .
\end{eqnarray}
Consider now two black hole solutions $(M,g)$ and $(\tilde M, \tilde g)$
as in the statement of the theorem.
We denote the corresponding matrices defined as above by $\Phi$ and
$\tilde \Phi$, and we use the same ``tilde'' notation to distinguish
any other quantities associated with the two solutions.
$\langle \! \langle M \rangle \! \rangle$ as a manifold with a $\G$-action
is uniquely determined by the interval structure modulo diffeomorphisms
preserving the action of $\G$ and similarly for the tilde spacetime.
Therefore, since the interval structures
are assumed to be the same for both spacetimes, $\langle \! \langle M \rangle \! \rangle$
and $\langle \! \langle \tilde M \rangle \! \rangle$
are isomorphic as manifolds with a $\G$ action, and we may hence
assume that $\tilde t^a = t^a,
\tilde \psi_i^a=\psi_i^a$ for $i=1, \dots, D-3$, and we may also assume
that $\tilde r= r$ and $\tilde z = z$.
Consequently, it is possible to combine the divergence
identities~\eqref{divergenceid} for the two solutions into a single
identity on the upper complex half plane, called ``Mazur identity''.
It is given by
\begin{eqnarray}\label{MazurId}
{\hat D}_{\alpha}(r {\hat D}^{\alpha} \sigma )
=
r \,
\hat g^{\alpha\beta} {\rm Tr} \, \left(\hat N^{T}_\alpha \hat N^{}_\beta \right) \, ,
\end{eqnarray}
and it can be proven in almost exactly the same way as the identity given in~\cite{Mazur}.
Here, we have written
\begin{equation}
\sigma = {\rm Tr} (\tilde \Phi \Phi^{-1}-I), \quad
\hat N_\alpha = \tilde S^{-1} (\tilde \Phi^{-1} \hat D_\alpha \tilde \Phi -
\Phi^{-1} \hat D_\alpha \Phi) S \, ,
\end{equation}
where in turn $S$ and $\tilde S$ are matrices such that
$\Phi = S^T S$ and $\tilde
\Phi = \tilde S^T \tilde S$ hold.
The key point about the Mazur identity~\eqref{MazurId}
is that on the left side we have a total divergence,
while the term on the right hand side is
non-negative. This structure can be exploited in various ways.
In this paper, we follow a strategy invented by Weinstein~\cite{Weinstein,Weinstein1},
which differs from that originally devised by Mazur.

The basic idea is to view $r,z$ as cylindrical coordinates in an auxiliary space
$\mr^3$ consisting of the points ${\bf x} = (r \cos \gamma, r \sin \gamma, z)$,
and to view $\sigma$ as a rotationally symmetric function defined on this
$\mr^3$, minus the $z$-axis. The Mazur identity then gives
\ben
\Delta \sigma \ge 0 \quad \text{on $\mr^3 \setminus \{z$-axis$\}$,}
\een
where $\Delta$ is the ordinary Laplacian on $\mr^3$. As we will show, $\sigma$ is
globally bounded on $\mr^3$, including at infinity and the $z$-axis. Furthermore, we claim that
$\sigma \ge 0$ at any point away from the axis: Writing $F = \tilde S S^{-1}$, we
have $\sigma = {\rm Tr} \, (F^T F) - (D-2)$. Now, $F^T F \ge 0$, and $\det \, F^T F = \det \tilde \Phi \, \det \Phi^{-1} = 1$,
so we may bring $F^T F$ into the form ${\rm diag}(\e^{u_1}, \dots, \e^{u_{D-3}}, \e^{-u_1- \dots - u_{D-3}})$
by a similarity transformation. Thus, $\sigma$ will be non-negative if and only if
\ben
\frac{1}{D-2} (\e^{u_1} + \dots + \e^{u_{D-3}} + \e^{-u_1- \dots - u_{D-3}}) \ge 1 \,,
\een
which in turn follows directly because the exponential function is convex.
Thus, we are in a position to apply the maximum principle arguments in~\cite{Weinstein1},
which imply that $\sigma = 0$
everywhere. As we now see, this implies that the metrics $g$
and $\tilde g$ are isometric on the domain of outer communication, thus proving
the theorem.

First, $\sigma = 0$ implies that $u_1 = \dots =  u_{D-3} = 0$, and hence that
$\tilde \Phi = \Phi$ everywhere in $\hat M$. Therefore,
the twist potentials  and the Gram matrices of the axial Killing fields are identical
for the two solutions, $\tilde f_{ij} = f_{ij}$ and $\tilde \chi_i =
\chi_i$. To see that
the other scalar products between the Killing fields coincide for the
two solutions, let $\alpha_i = g(t, \psi_i), \beta = g(t,t)$ as above, and define similarly the scalar products
$\tilde \alpha_i, \tilde \beta$ for the other spacetime.
The right side of eq.~\eqref{Dalpha} does not depend upon the conformal factor $\nu$, so
since $\tilde \chi_i = \chi_i$ and $\tilde f_{ij} = f_{ij}$, it also follows that
$\tilde \alpha_i = \alpha_i$ up to a constant. That constant has to vanish, since
it vanishes at infinity. Furthermore, from eq.~\eqref{alphabeta}
we have $\tilde \beta = \beta$.
Thus, all scalar products of the Killing
fields are equal for the two solutions, $\tilde G_{IJ} = G_{IJ}$ on the
entire upper half plane. Viewing now the second reduced Einstein
equation~\eqref{reduced3} as an equation for $\nu$
respectively $\tilde \nu$, and bearing in mind that $\nu = \tilde \nu$ at infinity, one
concludes that $\tilde \nu = \nu$. Thus, summarizing,
we have shown that if the boundary integral in the integrated
Mazur identity eq.~\eqref{MazurId} could be shown to vanish, then
$\tilde G_{IJ} = G_{IJ}$, $\tilde r = r$, $\tilde z = z$ and $\tilde \nu =
\nu$. Since $\tilde t = t, \tilde \psi_i = \psi_i$
it follows from eqs.~\eqref{metric} and~\eqref{hatg}
that $\tilde g = g$ in the domain of outer communication.

It remains to be shown that $\sigma$ is globally bounded, including
at the $z$-axis (corresponding to $\partial \hat M$) and at infinity.
It is here that the assumptions of the theorem
about the interval structures and angular momenta are needed.
We must consider the following separate cases: (1) The parts of
$\partial \hat M$ corresponding to a rotation axis of the Killing fields, (2)
the part corresponding to the horizon, and (3) infinity.
\begin{enumerate}
\item[(1)] On each segment $z \in I_j=(z_j, z_{j+1}), r=0$ of the
boundary $\partial \hat M$ representing an axis, we know
that the null spaces of the Gram matrices $f_{ij}$ and $\tilde f_{ij}$
coincide, because we are assuming
that the interval structures of both solutions are identical. Furthermore,
from eq.~\eqref{lineint}, and from the fact that $\hat \omega_i$
vanishes on any axis by definition, the twist potentials $\chi_i$
are constant on the $z$-axis outside of the segment $(z_h,z_{h+1})$
representing the horizon. The difference between the constant
value of $\chi_i$ on the $z$-axis left and right to the horizon
segment can be calculated as follows:
\begin{eqnarray*}
\chi_i(r=0,z_h) - \chi_i(r=0,z_{h+1})
&=&
\int_{z_h}^{z_{h+1}} \hat \omega_{i} \\
&=& \frac{1}{(2\pi)^{D-3}}
\int_{{\mathcal H}} *(\d\psi_i) \\
&=&
\frac{1}{(2\pi)^{D-3}}
\int_{\S^3 \times \T^{D-5}} *(\d\psi_i) =
\frac{1}{(2\pi)^{D-3}} \, J_i \, .
\end{eqnarray*}
The first equality follows from the definition of the twist
potentials, the second from the defining formula for the twist
1-forms and the fact that these are
invariant under the action of the $D-3$ independent rotation isometries
each with period $2\pi$ (with ${\mathcal H}$ a horizon cross section),
the third equation follows from Gauss' theorem and the fact that
$\d(*\d\psi_i)=0$ because $\psi_i$ is a Killing vector on a Ricci-flat manifold, and the last
equality follows from the Komar expression for the angular momentum.
The analogous expressions hold in the spacetime $(\tilde
M, \tilde g)$.
Because we assume that $J_i = \tilde J_i$, we can
add constants to $\chi_i$, if necessary, so that $\chi_i = \tilde
\chi_i$ on the axis, and in fact that $\chi_i - \tilde
\chi_i = O(r^2)$ near any axis.
One may now analyze the behavior of $\sigma$ near our boundary segment $I_j=(z_j, z_{j+1}), r=0$,
given by
\begin{equation}\label{difference}
\sigma
= -1+\frac{\det f}{\det \tilde f} + \frac{f^{ij} (\chi_i -\tilde \chi_i)(\chi_j-\tilde \chi_j)}{{\rm det} \, \tilde f}
+ f^{ij}(\tilde f_{ij}-f_{ij})  \, .
\end{equation}
Let $\underline a(I_j) \in \Z^{D-3}$ be the vector generating the kernel of the matrix
$f$ in on our interval $I_j$. By lemma~\ref{lemma2}, we can find a matrix $B \in SL(D-3, \Z)$
such that $\underline a(I_j) B^T= (1, 0, \dots, 0)$. Thus, redefining the
axial Killing fields to $\hat \psi_i = \sum_j A_{ij} \psi_j$ and $A = B^{-1}$ if necessary, we can assume
without loss of generality
that $\underline a(I_j) = (1, 0, \dots, 0)$. By arguments parallel to those in the proof
of lemma~\ref{lemma6}, the matrix $\hat f$ then takes the following form near $I_j$:
\ben
\hat f \sim \left(
\begin{matrix}
r^2 \e^{2\nu} & O(r^2) & \dots & O(r^2)\\
O(r^2) & \hat f_{22} & \dots & \hat f_{2(D-3)}\\
\vdots & \vdots & & \vdots \\
O(r^2) & \hat f_{(D-3)2} & \dots & \hat f_{(D-3)(D-3)}
\end{matrix}
\right)  \, ,
\een
and similarly for the second solution. It follows from this expression and
eq.~\eqref{difference} that $\sigma$ is finite near the interval $I_j$.
\item[(2)]
On the horizon segment, the matrices $f_{ij}, \tilde f_{ij}$ are
invertible, so $\sigma$ is finite there.
\item[(3)] Near infinity, the matrices $f_{ij}$ and $\tilde f_{ij}$ both tend to
the same limiting matrix, as is evident from our discussion of the asymptotic form
of the metric in sec.~\ref{sec3}.  Thus, $f_{ij}-\tilde f_{ij} = O(1/\rho)$, where
$\rho = \sqrt{r^2+z^2} = |{\bf x}|$.
We claim that the twist potentials $\chi_i$ and $\tilde \chi_i$ also approach the same value near infinity. First, we already know that they are equal on the axis $r=0$ and away from the horizon interval. We now take a path in eq.~\eqref{lineint} that stays on the axis $r=0$ until a very large $|z|$, and then
follows a half circle $\sqrt{r^2+z^2} = |{\bf x}| = const.$ in the asymptotic region. In this
region, we can use the asymptotic form of the metric, eq.~\eqref{metricinfty} derived above.
In the coordinates $\rho = |{\bf x}|$ and $\tan \theta = r/z$ as in eq.~\eqref{metricinfty}, the $\d\theta$-component of
the twist 1-forms are given to leading order in $1/\rho$ by
\ben
\hat \omega_i = \frac{1}{2}(2\pi)^{3-D} \, J_i
\begin{cases}
(1 -(-1)^i \cos \theta)\sin \theta \, \d\theta + \dots & \text{for $i=1,2$,}\\
%const. \, J_2 (1+\cos \theta) d\cos \theta + \dots & \text{for $i=2$,}\\
\sin \theta \, \d\theta + \dots & \text{for $i=3,\dots,D-3$,}
\end{cases}
\een
where dots stand for terms of higher order in $1/\rho$, or terms proportional to $\d\rho$, which
do not contribute in a line integral as in~\eqref{lineint} along a large circle of constant $\rho$.
The asymptotic behavior of the twist potentials is hence
\ben
\chi_i = \frac{1}{2}(2\pi)^{3-D} \, J_i
\begin{cases}
\cos \theta -(-1)^i \tfrac{1}{2} \cos^2 \theta-1+(-1)^i\tfrac{1}{2} + \dots & \text{for $i=1,2$,}\\
%const. \, J_2 (1+\cos \theta) d\cos \theta + \dots & \text{for $i=2$,}\\
\cos \theta -1+ \dots & \text{for $i=3,\dots,D-3$,}
\end{cases}
\een
where the dots stand for terms of order $O(1/\rho)$.
The same formula holds for the tilde quantities.
Therefore, because $J_i = \tilde J_i$,
it follows that $\chi_i - \tilde \chi_i = O(1/\rho)$. Thus,
$\sigma$ tends to a zero for $|{\bf x}| \to \infty$.
\end{enumerate}
Thus we have shown that $\sigma$ remains bounded, including the axis, horizon segment, and tends to
zero near infinity.
As we have explained, this concludes the proof of the theorem. \qed

\section{Conclusions and outlook}

In this paper, we have proved a uniqueness theorem for $D$-dimensional stationary, asymptotically Kaluza-Klein
black hole spacetimes satisfying the vacuum Einstein equations, allowing a group of
isometries $\G = \mr \times \T^{D-3}$. We showed that the solutions are uniquely determined by
certain combinatorial data specifying the group action, certain moduli, and the angular momenta.
This combinatorial data in particular determines the topology of the spacetime outside the black hole,
and the topology of the horizon.

To be able to prove our uniqueness theorem, we also had to make a number technical assumptions. They
mainly concern the analyticity of the metric and the causal structure of the spacetime.
One feels that it ought to be possible to remove these assumptions, but it is not clear to us
how this could be done in practice.

The more unsatisfactory aspect of our analysis is that we have not been able to prove or disprove the existence
of smooth black hole solutions associated with more elaborate topological structure/combinatorial data, such
as ``black lenses'' etc.
Some partial results have been obtained in the literature on this (see e.g.~\cite{chen}), but the general situation is still unclear.

\medskip
\noindent
{\bf Acknowledgements:} S.H. would like to thank Iskander~Aliev for discussions about lattices, and
Piotr Chrusciel for extensive discussions on manifolds with torus actions.
S.Y. gratefully acknowledges support by the Alexander von Humboldt
Foundation and the Sofia University Research Fund under grant No~111.
\appendix

\section{Proof of lemma~\ref{lemma6}}\label{appendixa}

{\bf Lemma~6:} The length of the horizon interval satisfies
\ben
(2\pi)^{D-3} l_H = \kappa A_H \,\,\,  ,
\een
where $A_H$ is the area of the horizon cross section $\H$, and where
$\kappa>0$ is the surface gravity.

\medskip
\noindent
{\em Proof:} We take the horizon to correspond to the interval $z \in (z_1, z_2), r=0$
on the boundary of the orbit space $\hat M$. Let $\underline v = (1, \Omega_1, \dots, \Omega_{D-3})$. Then by definition
$G_{IJ} v^I v^J = g(K, K)$, where $K$ is the Killing vector~\eqref{Kdef}, which
is tangent to the null generators of the horizon $H$, so $G_{IJ} v^I v^J = 0$ on $H$.
It the follows e.g. from the min-max principle that $G_{IJ}v^J = 0$ on the horizon, so $\lim_{r \to 0}
G_{IJ}v^J = 0$ in the orbit space for $z \in (z_1, z_2)$.
As was shown in~\cite[sec. 3]{Harmark}, one can furthermore use the first reduced Einstein equation~\eqref{reduced2} to
show that $\lim_{r \to 0} G_{IJ}v^J/r = 0$ for $z \in (z_1, z_2)$.

Let us now choose coordinates $(u,r,\varphi_1, \dots, \varphi_{D-3})$ near $H$ such that
$K = \partial/\partial u, \psi_i = \partial/\partial \varphi_i$. Let us
define $\hat X_I$ as $X_I$ above in eq.~\eqref{gramdef}, with $t$ replaced by $K$,
and let $\hat G_{IJ} = g(\hat X_I, \hat X_J)$. Then the reduced Einstein equations
also hold for $\hat G$, and furthermore, near $r=0$ and $z \in (z_1, z_2)$, we have
\ben
\hat G \sim \left(
\begin{matrix}
- r^2 {\rm det} f^{-1} & O(r^2) & \dots & O(r^2) \\
O(r^2) & f_{11} & \dots & f_{1(D-3)}\\
\vdots & \vdots & & \vdots \\
O(r^2) & f_{(D-3)1} & \dots & f_{(D-3)(D-3)}
\end{matrix}
\right)  \, ,
\een
up to terms of higher order in $r$. Here, $z \in (z_1, z_2)$, and $f_{ij}(z)$
is the limit as $r \to 0$ of $g(\psi_i, \psi_j)$. Following~\cite[sec. 3]{Harmark}, the
second reduced Einstein equation~\eqref{reduced3} furthermore gives
\ben
\partial_r \nu \to 0 \, , \quad \partial_z \nu \to -\frac{1}{2} \partial_z \log {\rm det} f \, ,
\quad \text{as $r \to 0, z \in (z_1, z_2)$.}
\een
We conclude from the last relation that
$\e^{-2\nu} \to c^2 {\rm det} \, f$ for some constant $c>0$ as $r \to 0, z\in (z_1, z_2)$. From the form
of the metric given in eq.~\eqref{metric} (with $G$ replaced by $\hat G$), it follows that, near
$H$, we have
\bena
g &\sim& \e^{2\nu}(\d z^2 + \d r^2 - c^2 r^2 \d u^2) + \sum_{i,j=1}^{D-3} f_{ij}(z) \, \d\varphi_i \d\varphi_j
+ 2r^2 \sum_{i=1}^{D-3} O(1) \, \d u \d \varphi_i  \nonumber\\
&=&   \e^{2\nu} (\d z^2 + \d U\d V) + \sum_{i,j=1}^{D-3} f_{ij}(z) \, \d\varphi_i \d\varphi_j
+ \frac{1}{c} \sum_{i=1}^{D-3} O(1) \,  (V\d U-U\d V) \d\varphi_i \, .
\eena
The minus sign in front of the $\d u^2$-term follows from the fact that $K$ is timelike in a neighborhood outside $H$,
which in turn follows directly from $\nabla_a (K^b K_b) = \kappa K_a$. In the last line we
switched to Kruskal-like coordinates $U,V$ defined by
$UV = r^2, U/V = e^{2cu}$. It is apparent in these coordinates that $H$ corresponds to $V=0$. The restriction of
$K=\partial/\partial u$ to $H$ is found to be $cU \, \partial/\partial U$, from which one concludes in view of the equation
$K^a \nabla_a K^b = \kappa K^b$ on $H$ that $c=\kappa$. The lemma may now be proven by calculating the horizon
area in the coordinates $z,\varphi_i$ using the above form of the metric. It is
\ben
A_H = \int_{z_1}^{z_2} \d z \left( \prod_i \int_0^{2\pi} \d\varphi_i \right) \sqrt{\e^{2\nu} {\rm det} \, f}
    = \frac{1}{\kappa} (2\pi)^{D-3} (z_2-z_1) \, ,
\een
from which the lemma follows immediately in view of $l_H = z_2-z_1$.
\qed


\begin{thebibliography}{99}
\bibitem{Bunting}
Bunting,~G.~L.:
Proof of the uniqueness conjecture for black holes,
(PhD Thesis, Univ. of New England, Armidale, N.S.W., 1983)

\bibitem{Carter}
Carter,~B.:
Axisymmetric black hole has only two degrees of freedom,
Phys. Rev. Lett. {\bf 26}, 331-333 (1971)

\bibitem{cassels}
Cassels, J.W.S.: ``An introduction to the geometry of numbers,''
Springer Grundlehren der Mathematischen Wissenschaften Bd.~99,
(1959)

\bibitem{chen}
Chen, Y., Teo, E.:  A rotating black lens solution in five dimensions,
Phys. Rev. {\bf D78}: 064062,2008; arXiv:0808.0587 [hep-th]

\bibitem{KK2}
Cho, Y. M. and Freund, P. G. O.:
Non-Abelian gauge fields as Nambu-Goldstone fields,
Phys. Rev. D {\bf 12}, 1711 (1975)

\bibitem{Chrusciel}
Chrusciel, P.T.: On rigidity of analytic black holes,
Commun. Math. Phys. {\bf 189}, 1-7 (1997)

\bibitem{chrcos}
 Chrusciel, P.~T. and Lopes Costa, J.:
  ``On uniqueness of stationary vacuum black holes,''
  arXiv:0806.0016 [gr-qc].

\bibitem{chrnew}
 Chrusciel, P. T.:
  ``On higher dimensional black holes with abelian isometry group,''
  arXiv:0812.3424 [gr-qc].


\bibitem{Chrusciel:2008uz}
 Chrusciel,  P.~T., Galloway, G. J. and Solis, D.:
  ``Topological censorship for Kaluza-Klein space-times,''
  arXiv:0808.3233 [gr-qc].

\bibitem{holchru}
Chrusciel, P. and Hollands, S.: ``Manifolds with cohomogeneity-2 actions
of the torus group,'' in preparation.

\bibitem{elvang}
Elvang, H. and Figueras, P.:
  ``Black Saturn,''
  JHEP {\bf 0705}, 050 (2007)
  [arXiv:hep-th/0701035].

\bibitem{elvang1}
%\bibitem{Elvang:2004iz}
  Elvang, H., Harmark, T. and Obers, N. A.:
  ``Sequences of bubbles and holes: New phases of Kaluza-Klein black holes,''
  JHEP {\bf 0501}, 003 (2005)
  [arXiv:hep-th/0407050].


\bibitem{Emparan}
Emparan,~R. and Reall,~H.~S.: A rotating black ring in five dimensions.
Phys.\ Rev.\ Lett.\  {\bf 88}, 101101 (2002)

\bibitem{Reall}
Emparan,~R. and Reall,~H.~S.: Generalized Weyl solutions, Phys. Rev. D
{\bf 65}, 084025 (2002)

\bibitem{evslin}
Evslin, J.: Geometric Engineering 5d Black Holes with Rod Diagrams,
JHEP {\bf 0809} 004, 2008; arXiv:0806.3389 [hep-th]

\bibitem{Friedrich}
Friedrich, H., Racz, I. and Wald, R.M.:
On the rigidity theorem for spacetimes with a stationary event horizon
or a compact Cauchy horizon,
Commun. Math. Phys. {\bf 204}, 691-707 (1999)

\bibitem{Woolgar}
Galloway,~G.~J., Schleich,~K.,~Witt,~D.~M., and Woolgar,~E.:
Topological censorship and higher genus black holes,
Phys.\ Rev.\ D {\bf 60}, 104039 (1999)

\bibitem{Woolgar1}
Galloway,~G.~J., Schleich,~K., Witt.~D., and Woolgar,~E.:
The AdS/CFT correspondence conjecture and topological censorship,
Phys.\ Lett.\ B {\bf 505}, 255 (2001)


%\bibitem{Galloway}
%Galloway, G.J., Schoen, R.: A Generalization of Hawking's
%black hole topology theorem to higher dimensions,
%Commun. Math. Phys. {\bf 266}, 571 (2006)

\bibitem{Gibbons}
Gibbons~G.~W., Ida,~D., and Shiromizu,~T.:
Uniqueness and non-uniqueness of static black holes in higher dimensions,
Phys.\ Rev.\ Lett.\  {\bf 89}, 041101 (2002)

\bibitem{Oelsen}
  Harmark~T. and Olesen,~P:
  On the structure of stationary and axisymmetric metrics,
  Phys.\ Rev.\  D {\bf 72}, 124017 (2005)
  [arXiv:hep-th/0508208].
  %%CITATION = PHRVA,D72,124017;%%

%\cite{Harmark:2004rm}
\bibitem{Harmark}
  Harmark, T.:
  Stationary and axisymmetric solutions of higher-dimensional general
  relativity,
  Phys.\ Rev.\  D {\bf 70}, 124002 (2004)
  [arXiv:hep-th/0408141].
  %%CITATION = PHRVA,D70,124002;%%

\bibitem{Harmarktalk}
See talk by Harmark, T. available at: $\small \rm http:/\!/online.itp.ucsb.edu/online/highdgr06\\
/harmark1/pdf/Harmark-KITP.pdf$

\bibitem{H72}
Hawking, S.W.:
Black holes in general relativity.
Commun. Math. Phys. {\bf 25}, 152-166 (1972)

\bibitem{Hawking}
Hawking, S.W. and Ellis, G.F.R.:
{\em The large scale structure of space-time}
Cambridge: Cambridge University Press, 1973

%\bibitem{Heusler}
%Heusler, M.: {\em Black hole uniqueness theorems,}
%Cambridge University Press (1996)

\bibitem{Hirzebruch}
Hirzebruch, F.: ``Differentiable manifolds and quadratic forms,''
Lect. Notes. Univ. of California, Berkely (1962)

\bibitem{HIW}
  Hollands, S., Ishibashi, A. and Wald, R. M.:
  A higher dimensional stationary rotating black hole must be
  axisymmetric,
  Commun.\ Math.\ Phys.\  {\bf 271}, 699 (2007)
  [arXiv:gr-qc/0605106].

  \bibitem{Hollands:2008wn}
  Hollands, S. and Ishibashi, A.
  ``On the `Stationary Implies Axisymmetric' Theorem for Extremal Black Holes
  in Higher Dimensions,''
  arXiv:0809.2659 [gr-qc].


%\bibitem{HI}
%  Hollands, S. and Ishibashi, A.:
%  Asymptotic flatness and Bondi energy in higher dimensional gravity,
%  J.\ Math.\ Phys.\  {\bf 46}, 022503 (2005)
%  [arXiv:gr-qc/0304054].

\bibitem{Hollands:2007aj}
  Hollands, S. and Yazadjiev, S.:
  ``Uniqueness theorem for 5-dimensional black holes with two axial Killing
  fields,''
  Commun.\ Math.\ Phys.\  {\bf 283}, 749 (2008)
  [arXiv:0707.2775 [gr-qc]].

\bibitem{Hollands:2007qf}
  Hollands S., and Yazadjiev, S.:
  ``A Uniqueness theorem for 5-dimensional Einstein-Maxwell black holes,''
  Class.\ Quant.\ Grav.\  {\bf 25}, 095010 (2008)
  [arXiv:0711.1722 [gr-qc]].



%\bibitem{Ishihara}
%Ishihara, H., Kimura, M., Masuno, K., Tomizawa, S.:
%Black holes on Euguchi-Hanson space in five-dimensional
%Einstein Maxwell theory, Phys. Rev. D {\bf 74}, 047501 (2006)

\bibitem{Israel67}
Israel,~W.:
Event horizons in static vacuum space-times,
Phys. Rev., {\bf 164}, 1776-1779 (1967)

\bibitem{traschen}
  Kastor, D., Ray, S. and Traschen, J.:
  The First Law for Boosted Kaluza-Klein Black Holes,
  JHEP {\bf 0706}, 026 (2007)
  [arXiv:0704.0729 [hep-th]].

\bibitem{KK1}
Kerner, R.:
Generalization of Kaluza-Klein theory for an
arbitrary non-abelian gauge group,
Ann. Inst. H. Poincare, {\bf 9}, 143 (1968)

\bibitem{Kobayashi}
Kobayshi, S., Nomizu, K.: {\em Foundations of Differential
Geometry I}, Wiley 1969

\bibitem{Maison}
Maison,~D.: Ehlers-Harrison-type Transformations for Jordan's
extended theory of graviation, Gen. Rel. Grav. {\bf 10}, 717 (1979)

\bibitem{Massey}
Massey, W. S.: ``Algebraic Topology: An Introduction,'' Springer (1977)

\bibitem{Ida}
Morisawa, Y., Ida, D.: A boundary value problem for five-dimensional
stationary black holes, Phys. Rev. D {\bf 69}, 124005 (2004)

\bibitem{Mazur}
Mazur,~P.~O.:
Proof of uniqueness of the Kerr-Newman black hole solution,
J. Phys. A, {\bf 15}, 3173-3180 (1982)

\bibitem{Moncrief}
Moncrief, V. and Isenberg, J.:
Symmetries of cosmological Cauchy horizons.
Commun. Math. Phys. {\bf 89}, 387-413 (1983)

\bibitem{Moncrief:2008mr}
  Moncrief, V. and Isenberg, J.:
  ``Symmetries of Higher Dimensional Black Holes,''
  Class.\ Quant.\ Grav.\  {\bf 25}, 195015 (2008)
  [arXiv:0805.1451 [gr-qc]].


\bibitem{Myers}
Myers, R.C. and Perry, M.J.:
Black holes in higher dimensional space-times.
Annals Phys. {\bf 172} 304 (1986)

\bibitem{orlik1}
Orlik, P. and Raymond, F.: ``Actions of the torus on 4-manifolds I,''
Transactions of the AMS {\bf 152}, (1972)

\bibitem{orlik2}
Orlik, P. and Raymond, F.: ``Actions of the torus on 4-manifolds II,''
Topology {\bf 13} 89-112 (1974)

\bibitem{senkov}
  Pomeransky, A.A. and Sen'kov, R.A.:
  ``Black ring with two angular momenta,''
  arXiv:hep-th/0612005.


\bibitem{Racz}
Racz,~I.:
On further generalization of the rigidity theorem for spacetimes
with a stationary event horizon or a compact Cauchy horizon.
Class.\ Quant.\ Grav.\  {\bf 17}, 153 (2000)

\bibitem{Robinson}
Robinson,~D.~C.:
Uniqueness of the Kerr black hole,
Phys. Rev. Lett. {\bf 34}, 905-906 (1975)

\bibitem{Rogatko}
  Rogatko, M.:
  ``Uniqueness theorem of static degenerate and non-degenerate charged  black
  holes in higher dimensions,''
  Phys.\ Rev.\  D {\bf 67}, 084025 (2003)
  [arXiv:hep-th/0302091];
  ``Classification of static charged black holes in higher dimensions,''
  Phys.\ Rev.\  D {\bf 73}, 124027 (2006)
  [arXiv:hep-th/0606116].


\bibitem{Sudarsky}
Sudarsky, D. and Wald, R.M.:
Extrema of mass, stationarity, and staticity, and solutions to
the Einstein Yang-Mills equations.
Phys. Rev. D {\bf 46} 1453-1474 (1992)

\bibitem{Waldbook}
Wald, R.M.: {\em General Relativity}. Chicago:
University of Chicago Press, 1984

\bibitem{Weinstein}
Weinstein, G.: On rotating black holes in
equilibrium in general relativity,
Commun. Pure Appl. Math. {\bf 43}, 903 (1990)

\bibitem{Weinstein1}
See Lemma 8 in: Weinstein, G.: "On the Dirichlet
problem for harmonic maps with prescribed singularities,"
Duke Math. J. {\bf 77} (1995) No.1, 135-165
\end{thebibliography}
\end{document}